\documentclass[iop,apj,useAMS,usenatbib]{emulateapj-rtx4}

\usepackage{graphicx}
\usepackage{lscape}
\usepackage{amsmath}
\usepackage{longtable,natbib}
\usepackage[usenames,dvipsnames]{color}
\usepackage{gensymb}
\usepackage{multirow}

\usepackage{float}

\usepackage{amssymb} 
\usepackage[utf8]{inputenc} 
\usepackage{booktabs}
\usepackage{graphicx}

\journalinfo{Accepted to the Astrophysical Journal}
\submitted{}

\begin{document}


\title{Candidate gravitationally lensed dusty star-forming galaxies in the {\it Herschel}$^*$ Wide Area Surveys}
\author{H. Nayyeri \altaffilmark{1}, M. Keele \altaffilmark{1}, A. Cooray \altaffilmark{1}, D. A. Riechers \altaffilmark{2}, R.\,J.~Ivison \altaffilmark{3,4}, A.I. Harris \altaffilmark{5}, D. T. Frayer \altaffilmark{6}, A. J. Baker \altaffilmark{7}, S. C. Chapman \altaffilmark{8}, S. Eales \altaffilmark{9}, D. Farrah \altaffilmark{10}, H. Fu \altaffilmark{11}, L. Marchetti \altaffilmark{12}, R. Marques-Chaves \altaffilmark{13,14}, P. I. Martinez-Navajas \altaffilmark{13,14}, S. J. Oliver \altaffilmark{15}, A. Omont \altaffilmark{16}, I. Perez-Fournon \altaffilmark{13,14}, D. Scott \altaffilmark{17}, M. Vaccari \altaffilmark{18,19}, J. Vieira \altaffilmark{20},  M. Viero \altaffilmark{21}, L. Wang \altaffilmark{22},  J. Wardlow \altaffilmark{23,24}}

\altaffiltext{$\star$}{{\it Herschel} is an ESA space observatory with science instruments provided by European-led Principal Investigator consortia and with important participation from NASA.}
\altaffiltext{1}{Department of Physics \& Astronomy, University of California, Irvine, CA 92697}
\altaffiltext{2}{Department of Astronomy, Cornell University, Ithaca, NY, 14853}
\altaffiltext{3}{European Southern Observatory, Karl-Schwarzschild-Strasse 2, 85748 Garching, Germany}
\altaffiltext{4}{Institute for Astronomy, University of Edinburgh, Blackford Hill, Edinburgh EH9 3HJ, UK}
\altaffiltext{5}{Department of Astronomy University of Maryland College Park, MD 20742}
\altaffiltext{6}{National Radio Astronomy Observatory, Green Bank, WV, 24944}
\altaffiltext{7}{Department of Physics \& Astronomy, Rutgers, the State University of New Jersey, 136 Frelinghuysen Road, Piscataway, NJ 08854-8019}
\altaffiltext{8}{Institute of Astronomy, University of Cambridge, Madingley Road, Cambridge, CB3 0HA, U.K.}
\altaffiltext{9}{School of Physics \& Astronomy, Cardiff University, Cardiff, UK}
\altaffiltext{10}{Department of Physics, Virginia Polytechnic Institute and State University, Blacksburg, VA 24061}
\altaffiltext{11}{Department of Physics \& Astronomy, University of Iowa, Iowa City, Iowa 52242}
\altaffiltext{12}{Department of Physical Sciences, The Open University, Milton Keynes, MK7 6AA, UK}
\altaffiltext{13}{Instituto de Astrofisica de Canarias, E-38205 La Laguna, Tenerife, Spain}
\altaffiltext{14}{Universidad de La Laguna, Dpto. Astrofisica, E-38206 La Laguna, Tenerife, Spain}
\altaffiltext{15}{Department of Physics \& Astronomy, University of Sussex, Brighton BN1 9QH, UK}
\altaffiltext{16}{Institut d'Astrophysique de Paris, UMR 7095 CNRS, Universit\'e Pierre et Marie Curie, 75014, Paris, France}
\altaffiltext{17}{Department of Physics \& Astronomy, University of British Columbia, 2329 West Mall, Vancouver, BC V6T 1Z4, Canada}
\altaffiltext{18}{Department of Physics and Astronomy, University of the Western Cape, Robert Sobukwe Road, 7535 Bellville, Cape Town, South Africa}
\altaffiltext{19}{INAF - Istituto di Radioastronomia, via Gobetti 101, 40129 Bologna, Italy}
\altaffiltext{20}{Department of Astronomy and Department of Physics, University of Illinois, 1002 West Green St., Urbana, IL 61801}
\altaffiltext{21}{Department of Physics, Stanford University, Stanford, California 94305}
\altaffiltext{22}{SRON Netherlands Institute for Space Research, Landleven 12, 9747 AD, Groningen, The Netherlands}
\altaffiltext{23}{Dark Cosmology Centre, Niels Bohr Institute, University of Copenhagen, DK-2100 Copenhagen, Denmark}
\altaffiltext{24}{Centre for Extragalactic Astronomy, Department of Physics, Durham University, South Road, Durham, DH1 3LE, UK}


\begin{abstract}
We present a list of candidate gravitationally lensed dusty star-forming galaxies (DSFGs) from the HerMES Large Mode Survey (HeLMS) and the {\it Herschel} Stripe 82 
Survey (HerS). Together, these partially overlapping surveys cover 372\,deg$^{2}$ on the sky. 
After removing local spiral galaxies and known radio-loud blazars, our candidate list of lensed DSFGs 
is composed of 77 sources with 500\,$\mu$m flux densities ($S_{500}$) greater than 100\,mJy. Such sources are dusty starburst galaxies similar to the first bright Sub Millimeter Galaxies (SMGs) discovered with SCUBA. We expect a large fraction of this list to be strongly lensed, with a small fraction made up of bright SMG-SMG mergers that appear as Hyper-Luminous Infrared Galaxies (HyLIRGs; $\rm L_{IR}>10^{13}\,L_{\odot}$). Thirteen of the 77 candidates have spectroscopic redshifts from CO spectroscopy with ground-based interferometers, putting them at $z>1$ and well above the redshift of the foreground lensing galaxies. The surface density of our sample is 0.21 $\pm$ 0.03\,deg$^{-2}$. We present follow-up imaging of a few of the candidates confirming their lensing nature. The sample presented here is an ideal tool for higher resolution imaging and spectroscopic observations to understand detailed properties of starburst phenomena in distant galaxies.
\end{abstract}


\section{Introduction}

Dusty Star-Forming Galaxies (DSFGs) are among the most intensely star forming systems in the Universe (see review by \citealp{Casey2014}). Optical studies of these galaxies are challenging, due to high dust obscuration that absorbs the rest-frame ultraviolet (UV) light emitted by young and hot stars. Instead these sources are bright at the longer wavelengths due to the emission from heated dust \citep{Draine2001, Siebenmorgen2014}. The far-infrared luminous DSFGs are bright in the sub-mm wavelengths and are similar to the first bright Sub Millimeter Galaxies (SMGs) discovered with SCUBA \citep{Smail1997, Barger1998, Blain1999, Dunne2000, Chapman2005, Tacconi2006, Tacconi2008, Magnelli2012, Hayward2013, Swinbank2014, Wiklind2014, Ikarashi2015} with the brightest sources in the far-infrared bands (with infrared luminosities of $\rm 10^{12}\,L_{\odot}<L_{IR}<10^{13}\,L_{\odot}$) classified as Ultra-Luminous Infrared Galaxies (ULIRGs; \citealp{Sanders1988, Sanders1996, Lutz1999, Alonso2006, Clements2010, Kilerci2014, Magdis2014}). Resolved imaging of DSFGs at high redshifts is challenging, given the intrinsic faintness of such systems (with intrinsic flux densities of typically 10\,mJy at 500\,$\mu$m, from models or direct observations such as \citealp{Ivison2010, Bethermin2012a, Bethermin2012b} or from lensing magnification corrected 500\,$\mu$m studies such as \citealp{Negrello2010, Wardlow2013}) and/or large point spread functions of single dish diffraction-limited observations at sub-mm wavelengths. Although the latter is no longer a limitation, because of multi-dish observations such as with ALMA using long baselines \citep{Karim2013, Wang2013, Riechers2014, Simpson2015, ALMA2015}, the field of view remains limited to a few arcseconds \citep{Karim2013}. As a result, current studies of high-redshift DSFGs are generally limited to a small number of targets \citep{Capak2008, Riechers2010, Riechers2013, Fu2013, Riechers2014} and the most highly magnified star-forming systems (such as the Cosmic Eyelash; \citealp{Swinbank2010}) or extreme starbursts \citep{Coppin2009, Riechers2013, Gilli2014, DeBreuck2014, Riechers2014}.

Gravitational lensing provides a valuable tool to study galaxies that would otherwise be too distant or faint for current observational facilities (eg. \citealp {Treu2010, Treu2014}). This is due to the fact that lensing enhances apparent angular size and magnifies source flux density \citep {Richard2014, Atek2014},
enabling studies of sources with intrinsic brightness below the nominal source detection limits of current facilities \citep{Wardlow2013}. By searching blank-field sub-millimeter surveys for bright sources, we can select galaxies that have a higher probability to be magnified by lensing \citep{Swinbank2014, Simpson2015, Simpson2015b}. With the advent of large area far-infrared and sub-mm surveys it is now possible to search for bright sources as candidate gravitationally lensed systems with the selection based on the sub-mm flux density only \citep{Negrello2007, Negrello2010}.

The {\it Herschel} Space Observatory \citep{Pilbratt2010} provided us with a unique opportunity to study DSFGs at high redshift. This is possible through both large-area sky surveys, such as the {\it Herschel} Astrophysical Terahertz Large-Area Survey (H-ATLAS; \citealp{Eales2010}), and deeper observations but over smaller areas, such as those of the {\it Herschel} Multi-tiered Extragalactic Survey (HerMES; \citealp{Oliver2012}). The large area searches for gravitationally lensed systems has been very successful over the past few years at identifying high redshift DSFGs. In particular {\it Herschel} has been successful at identifying rare lensing systems, with some detailed studies already in the literature \citep{Ivison2010, Cox2011, Fu2012, Messias2014}. 
Follow-up observations of these candidates with ground-based facilities (such as ALMA; \citealt{Messias2014, Schaerer2015}) and the {\it Hubble} Space Telescope in the near-infrared has revealed the nature of the ISM in these systems with spatial resolutions at the level of 100 pc scales \citep{Karim2013, Swinbank2014, Riechers2014, Swinbank2015}. This allows us to study gas regulations and kinematics inside distant galaxies which sheds light on star formation mechanism and efficiency in the most gas-rich systems during the peak epoch of star formation in the Universe \citep {Riechers2014}. An interesting example is SDP.81, which was initially identified with {\it Herschel} during the Science Demonstration Phase (SDP; \citealp{Negrello2010})
as a lensed DSFG \citep{Frayer2011, Hopwood2011, Negrello2014, Dye2014}. SDP.81 was later used for Science Verification observations of ALMA long baselines \citep{ALMA2015b}. Those data, collected for more than 30 hours, have now resulted in a robust lens model and have enabled ISM studies by revealing clumpy structures and giant molecular clouds within the lensed galaxy down to 80 pc scales \citep{Dye2015, Wong2015, Swinbank2015, Rybak2015a, Rybak2015b, Hatsukade2015, Hezaveh2016}.

The aim of this work is to expand the currently known samples of bright, lensed galaxies from {\it Herschel}.
The first study of this kind used the Science Demonstration Phase map of H-ATLAS spanning 14\,deg$^2$ in \citet{Negrello2010}, identifying five candidate lenses that were confirmed to be lensed with follow-up data. The second systematic search for lensing systems in {\it Herschel} imaging data 
appeared in \citet{Wardlow2013}, selecting 13 candidate lensed systems over 95\,deg$^2$ of HerMES, composed of many smaller fields with individual sizes at the level of 2-10\,deg$^2$. Among those 13, eleven are now confirmed as strong lens systems; the two remaining systems are luminous SMG-SMG mergers \citep{Fu2013, Bussmann2015}.

Here we extend these two earlier studies with {\it Herschel}/SPIRE (Spectral and Photometric Imaging Receiver; \citealp{Griffin2010}) observations at 250, 350 and 500\,$\mu$m to 
select potentially gravitationally lensed galaxies in the HerMES Large Mode Survey (HeLMS) and {\it Herschel} Stripe 82 Survey (HerS; \citealp{Viero2014}) fields. The {\it Herschel} HeLMS is a wide, SPIRE-only observations covering an area of 301\,deg$^2$ in the SDSS Stripe 82 region with ancillary data from several facilities (\citealp{Oliver2012}, Clarke et al. 2016). The SPIRE observations reach a 5$\sigma$ limiting depth of 48\,mJy in the 500\,$\mu$m. The equatorial SDSS Stripe 82 HerS observations cover an area of 81\,deg$^2$ with an average depth of 14.8\,mJy beam$^{-1}$ (\citealp{Viero2014}, Clarke et al. 2016). This along with the ancillary data available in these equatorial fields, in particular deep SDSS observations from Stripe 82, enable us to identify and study these lensed galaxies. 

The catalog presented here has {\it Herschel} photometry measured in the three SPIRE bands. Follow-up observations with the Keck/NIRC2 using Laser-guided adaptive optics (LGS-AO) and with the seeing-limited William Herschel Telescope (WHT) LIRIS near-IR imaging instrument reveal the lensing nature of several of these systems. We also present follow-up observations of some of these sources with mm-wave interferometric spectroscopy of CO emission lines to determine the redshift of background lensed galaxies. In particular, we present redshifts measured for targeted background lensed DSFGs with CO\,($1 \rightarrow 0$) observations by the Green Bank Telescope (GBT)/Zpectrometer (Harris et al., in prep) and with multiple CO lines from Combined Array for Research in Millimeter-wave Astronomy (CARMA) and Plateau de Bure Interferometer (PdBI) (Riechers et al., in prep) showing that the confirmed lensed galaxies are at $z >1$ with a target being at redshift as high as $z \sim 5$. The SDSS-detected foreground lensing galaxies have mean photometric redshifts peaking at $z \sim 0.4$ with spectroscopic observations for some of the targets confirming $z<1$.

The paper is organized as follows. In Section 2, we discuss the lensed galaxy selection along with photometry measurements. Section 3 describes some of the follow-up programs and existing results. 
We discuss our candidate lens sample, statistical properties and some example lensed sources in Section 4.  
We conclude with a summary in Section 5. Throughout this paper we assume a standard cosmology with ${\rm H_0}=70\:\text{kms}^{-1}\text{Mpc}^{-1}$, $\Omega_m=0.3$ and
$\Omega_\Lambda=0.7$.

\section{Identification of Candidate Sources}

Our candidate lensed DSFGs are selected from HeLMS and HerS blank field data\setcounter{footnote}{0}\footnote{http://www.astro.caltech.edu/hers/Science.html}. HeLMS and HerS cover 301.3 and 80.5\,$\rm deg^{2}$ respectively, overlapping in a $\sim$\,10\,deg$^{2}$ region. HeLMS is the widest area tier of HerMES \citep{Oliver2012}, the SPIRE Team's Guaranteed Time Observations (GTO) with {\it Herschel} Space Observatory \citep {Pilbratt2010}. We use the maximum likelihood mapmaker {\sc sanepic} (Signal and Noise Estimation Procedure Including Correlations; \citealp{Patanchon2008}) to create our maps. We refer the reader to \citet{Asboth2016} and Clarke et al. (2016) for details 
related to HeLMS, data processing, and map making. Similar details related to HerS are available in \citet{Viero2014}. The maps we produced are comparable in quality to publicly available maps from HerS, while for the HeLMS data we compare our maps to the results from the SMAP/SHIM iterative map maker \citep {Levenson2010}. The differences, if any, are at large angular scales related to the diffuse background. For point sources and point source flux estimation, the different maps showed a comparable performance within the uncertainties related to instrumental and confusion noise. The nominal pixel sizes at 250, 350 and 500\,$\mu$m are 6, 8.3 and 12 arcseconds, respectively, matching one third of the full-width half-maximum (FWHM) of the beam in each band (18, 25, and 36 arcseconds respectively). We used and compared multiple catalogs as part of this study. The first set of catalogs make use of {\sc SUSSEXtractor} \citep{Savage2007, Smith2012}, available from the {\it Herschel} Interactive Processing Environment (HIPE; \citealp{Ott2010}), based on sources detected at each of 250, 350 and 500\,$\mu$m. The second set is discussed in Clarke et al. (2016) using {\sc starfinder} and {\sc desphot}. The latter makes use of 250\,$\mu$m detections to cross-identify and de-blend at 350 and 500\,$\mu$m flux densities, which reduces contamination from blended sources (XID catalogs). 

Here we primarily focus on bright sources with flux densities above 100\,mJy at 500\,$\mu$m. While our initial selection is with catalogs based on each of the wavelengths, using {\sc SUSSEXtractor} catalogs, the final selection involves removal of sources that are flagged as potentially blended systems using XID catalogs of Clarke et al. (2016). Unlike Clarke et al. (2016), which requires a 250\,$\mu$m detection, we are able to
search for and catalog lensed candidates that are also ``red'' with $S_{\rm 250} < S_{\rm 350} <  S_{\rm 500}$ \citep{Asboth2016}. Bright 500\,$\mu$m sources are dominated by gravitationally lensed galaxies, local spiral galaxies, and blazars \citep{Negrello2010, Wardlow2013}. Both local spirals and blazars can be excluded via cross-matching with shallow full sky surveys at optical and radio wavelengths, respectively. This method has been exploited recently in other {\it Herschel} fields to select high-redshift gravitationally lensed galaxies \citep{Negrello2010, Conley2011, Gonzalez2012, Bussmann2013,Wardlow2013, Negrello2014}. After correction for contamination, this technique has been shown to be very successful at generating a robust catalog of gravitationally lensed high redshift galaxies \citep{Negrello2010}. 

We remove local galaxies from our catalogs by searching the NASA/IPAC Extragalactic Database (NED) and the Sloan Digital Sky Survey III Data Release 12 at the 250\,$\mu$m positions of our sources. Local galaxies are then recognized by visual inspection of the SDSS cutouts and identified by name in NED. There are 273 such sources in our fields, 231 from HeLMS and 44 from HerS (two lie in the overlapping region). These sources are all within 200\,Mpc. As these local galaxies are rare (0.74\,deg$^{-2}$) and occupy only a small portion of the fields, positional alignment between background dusty star-forming galaxies and these local galaxies should not be significant. Furthermore, the typical distance ratio between foreground and background population is such that, in the case of chance alignments, the foreground galaxy will not act as a strong gravitational lens. Hence we do not expect to lose bona fide strongly lensed galaxies by removing the local spirals. The relative colors of the local galaxies and lens candidates also typically occupy different regions of the color space. 
We also remove radio-loud blazars in a similar fashion, searching NED for all sources within 12 arcseconds of the position of the peak 500\,$\mu$m flux. Sources identified as radio-loud quasars in previous surveys are then removed from our target list. We found nine blazars using this method. Note that our search was simply limited to known radio-loud blazars and we did not search for new blazar candidates. Given the existence of archival radio surveys down to sufficient depth we do not consider contamination from unknown blazars to be a significant issue with the candidate lensed sample.

After removal of the aforementioned contaminants, we examine our target lists from HeLMS and HerS for sources present in both catalogs from the 15\,deg$^{2}$ region covered by both surveys. These duplicate sources are found by matching every source in the HeLMS catalog with the source with nearest position in HerS, as measured by peak $S_{250}$ flux. The SPIRE observations at the 250\,$\mu$m have better angular resolution (FWHM\,=\,18$^{\prime\prime}$) compared to the redder bands which makes it more suitable for the cross-matching. Duplicate sources have a nominal separation that is small compared to the separation between sources that are merely nearby in a two-dimensional sky projection, between 1.3 and 7.2 arcseconds in all cases. In comparison, the closest two distinct sources in the combined catalog are separated by 45 arcseconds. The flux densities at 250, 350, and 500\,$\mu$m, and the colors of the sources, can also be compared to provide confirmation that sources from the two catalogs are indeed duplicates. Through this process, we found eight lens candidates that are present in both HeLMS and HerS, as well as the two local galaxies mentioned previously. In these cases, we remove the source from the HeLMS catalog and use the position and flux data from the deeper HerS data in all catalogs and analysis. 

We next generate cut-outs at the positions of the peak $S_{250}$ flux from the HeLMS and HerS maps at 250, 350, and 500\,$\mu$m for all sources with $S_{500}>$\,100\,mJy, which together define our primary lens candidate catalog. These cutouts are presented in the Appendix. From this list, we removed any remaining spatially extended sources and sources that are blended in $S_{500}$ compared to the 250\,$\rm \mu m$ imaging. Extended sources that are bright at sub-millimeter wavelengths, but are not local galaxies, are primarily galactic cirrus clouds and are not of interest for this catalog \citep{Low1984, Silva1998, Rowan2014}. We identify point sources that are separate by at least two beam sizes (separations $\sim 40^{\prime\prime}$) in the {\it Herschel}/SPIRE 250\,$\mu$m band. This should include any cluster lensed galaxies such as the HLSW-01 with a separation of $9^{\prime\prime}$, identified in the HerMES \citep{Scott2011, Conley2011, Riechers2011, Gavazzi2011}. We further cross matched our catalog of HeLMS/HerS sources with $S_{500}>$\,100\,mJy against the photometrically selected galaxy cluster catalog of \citet{Durret2015} in the Stripe 82. We found no cluster lensing systems within our 500\,$\mu$m bright sources.

As mentioned above one of the caveats of the {\sc SUSSEXtractor} catalogs is the difficulty in de-blending sources. In order to check the robustness of our bright 500\,$\rm \mu m$ lensed DSFG candidates we cross-checked our {\sc SUSSEXtractor} identified sources with the {\sc starfinder}+250\,$\rm \mu m$-detected XID catalog (\citealp{Viero2014}, Clarke et al. 2016), which used 250\,$\mu$m positions to de-blend 500\,$\mu$m flux densities. The only disadvantage of this catalog is that it relies on 250\,$\mu$m detections and thus does not contain 500\,$\mu$m peakers with faint 250\,$\mu$m emission. Using this combination of two catalogs we identified close to 30 sources that appear to be clear blends in 250\,$\mu$m, but appear as a single source at 500\,$\mu$m. Since they are likely multiple faint sources, rather than a single bright object, we do not consider such cases to be reliable candidates for gravitationally lensed sources. This approach, however, also results in a potential removal of rare lensed galaxies with image separations at the level of 30 arcseconds or more. Such lensing will involve most massive galaxy clusters in the foreground and are best searched for by combining SPIRE catalogs and known galaxy cluster positions. Given that our goal
is to obtain a reliable list with high efficiency for lensing, and not necessarily a complete list of all lensed galaxies, we consider our approach to be adequate to increase the reliability that a higher fraction of the sources in our candidate list are gravitationally lensed DSFGs.

After removing blended sources, and sources that are extended - and thus like to be contaminants from Galactic cirrus clumps (see Clarke et al. 2016) - we are left with a total of 77 lensed candidate galaxies 
with $S_{500}>100$\,mJy in our primary candidate list. The $S_{500}>100$\,mJy candidates have surface density $0.21 \pm 0.03$\,deg$^{-2}$ over the HeLMS and HerS fields. Using a sample of lensed DSFGs in the HerMES field \citet{Wardlow2013} measured a space density of 0.14\,deg$^{-2}$ for systems with $S_{500}>$\,100\,mJy for a candidate lensed sample composed of 13 systems. Among the 13, two has since been identified to be SMG-SMG mergers, though in both cases, there is evidence for moderate lensing with magnification factors between 1.3 to 1.8 \citep{Fu2013, Bussmann2015}. With those two removed, the actual surface density of lensed galaxies with $S_{500}>$\,100\,mJy in HerMES is around 0.11\, deg$^{-2}$.
This is substantially smaller than the surface density of our candidate list at 0.21\, deg$^{-2}$. We expect that $\sim$\,15\% of our sample to be also composed of SMG-SMG mergers, similar to sources studied in \citet {Fu2013, Ivison2013}. The real difference is likely due to cosmic variance. The study by \citet{Wardlow2013} involves multiple smaller legacy fields that were used in HerMES to form 95\,deg$^2$. These extragalactic legacy fields, such as XMM-LSS or Bo\"otes, are carefully selected to be devoid of known low-$z$ ($z=0.1-0.3$) massive galaxy clusters or galaxy over-densities that could potentially lens background DSFGs. Thus, these fields may, on average, provide a lower optical depth to lensing than a blank sky field. This is also visible in comparison to another result. The first lens selection with {\it Herschel}/SPIRE in H-ATLAS resulted in five confirmed lensed galaxies
with a sky surface density of 0.35\,deg$^{-2}$. While this area was 14\,deg$^2$ it shows that large field-to-field variation in the number of lensed sources. This cosmic variance
is a combination of spatial distribution of massive foreground galaxies or galaxy over-densities and the clustering of background DSFGs. With proper statistics on the lensed fraction from
HeLMS and HerS, as well as the lensed sample over 650\,deg$^2$ of H-ATLAS that is yet to fully appear in the literature, we expect it will be possible to observationally constrain
the lensing optical depth variations of {\it Herschel}-selected DSFGs. A theoretical calculation of the expected cosmic variance will also be useful to address which parameters related to either the background DSFG population or foreground lenses can be constrained with such statistics.

\begin{figure}
\centering
\includegraphics[trim=2cm 0cm 0cm 0cm, scale=0.45]{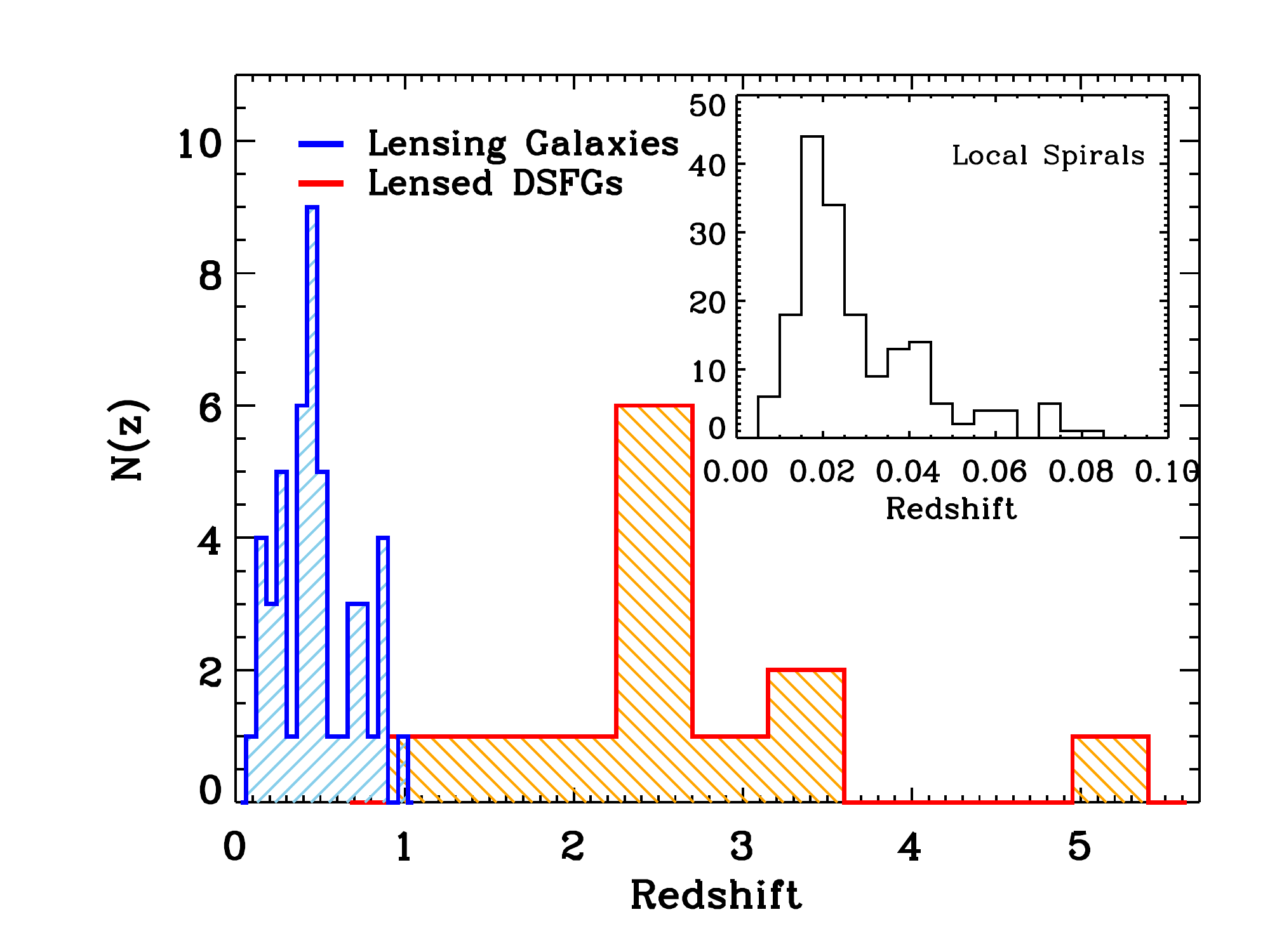}
\caption{Redshift distribution of the thirteen HeLMS and HerS candidate lensed galaxies with spectroscopic redshifts from CO observations (Harris et al., in prep, Riechers et al., in prep). The DSFGs are at a median redshift of 2.51, while the foreground galaxies have median redshift of 0.44 putting the background lensed systems at much higher redshifts compared to the foreground lensing galaxies. The inset shows the redshift distribution of the local spiral galaxies that are bright in the 500\,$\mu$m.}
\end{figure}

\begin{figure*}
\centering
\includegraphics[trim=2.5cm 1cm 0cm 12cm, scale=0.95]{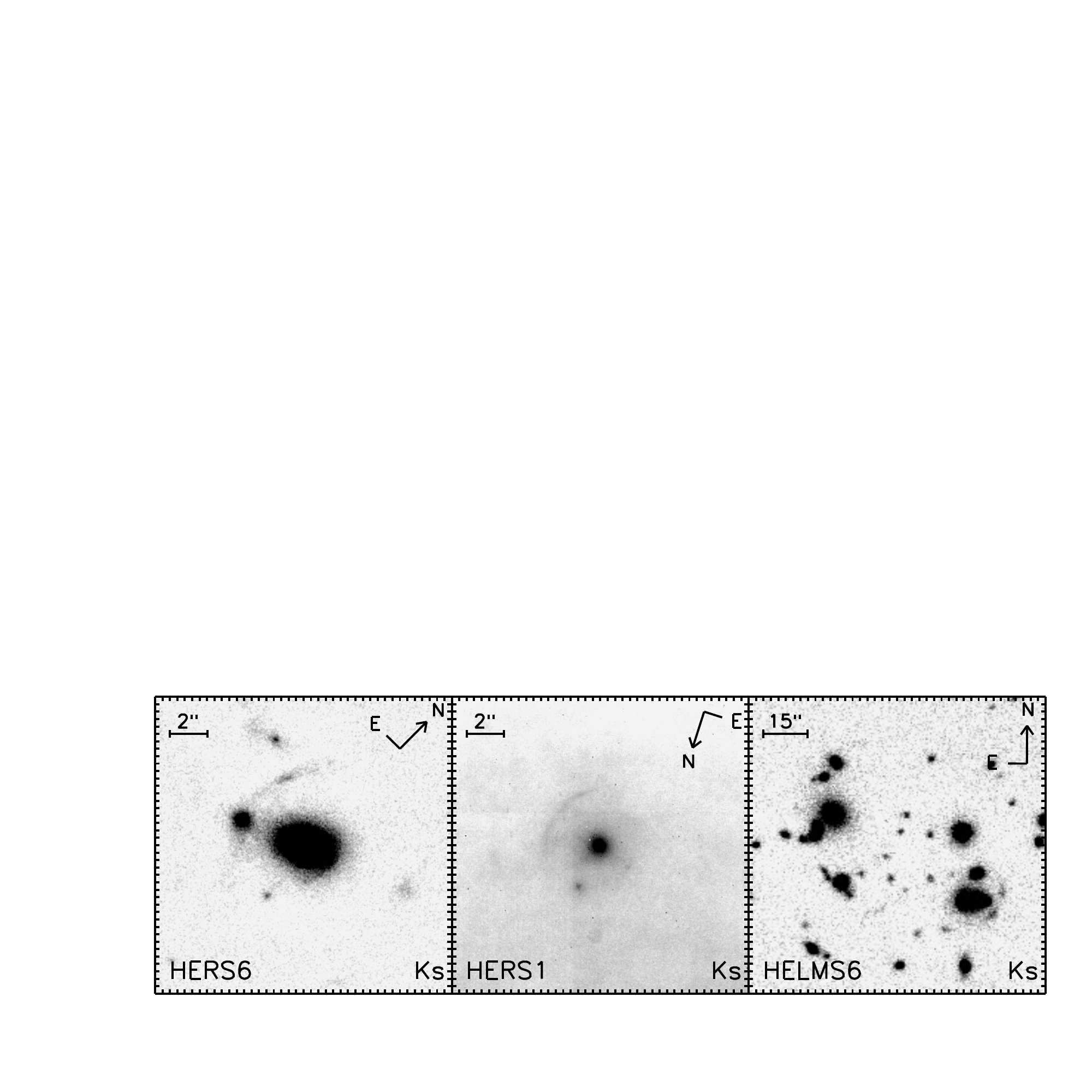}
\caption{Keck/NIRC2 AO observations of two of our lensed DSFG candidates (HERS1 and HERS6) in the $K_s$ band at 2.2\,$\mu$m along with William Herschel Telescope (WHT) LIRIS observations of HELMS6 in the $K_s$ band. We clearly see lensing features in these three systems that were originally identified in the SPIRE 500\,$\mu$m band. HERS1 is the brightest source in our catalog at 500\,$\mu$m and it has been confirmed to be gravitationally lensed in a previous study by \citet{Geach2015} and recently with the Atacama Cosmology Telescope \citep{Su2015}.}
\end{figure*}

\begin{figure}
\centering
\includegraphics[scale=0.4]{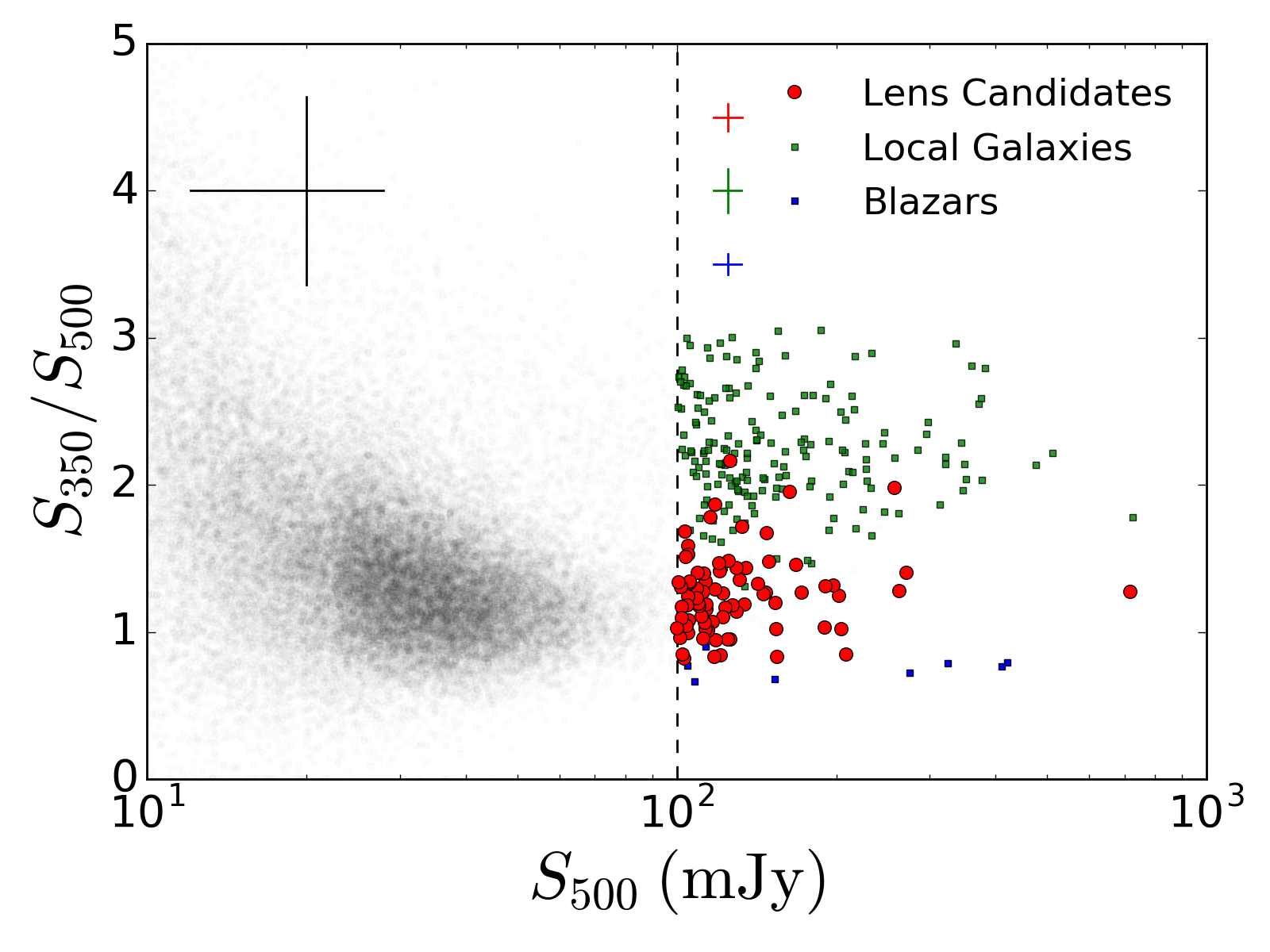}
\includegraphics[scale=0.4]{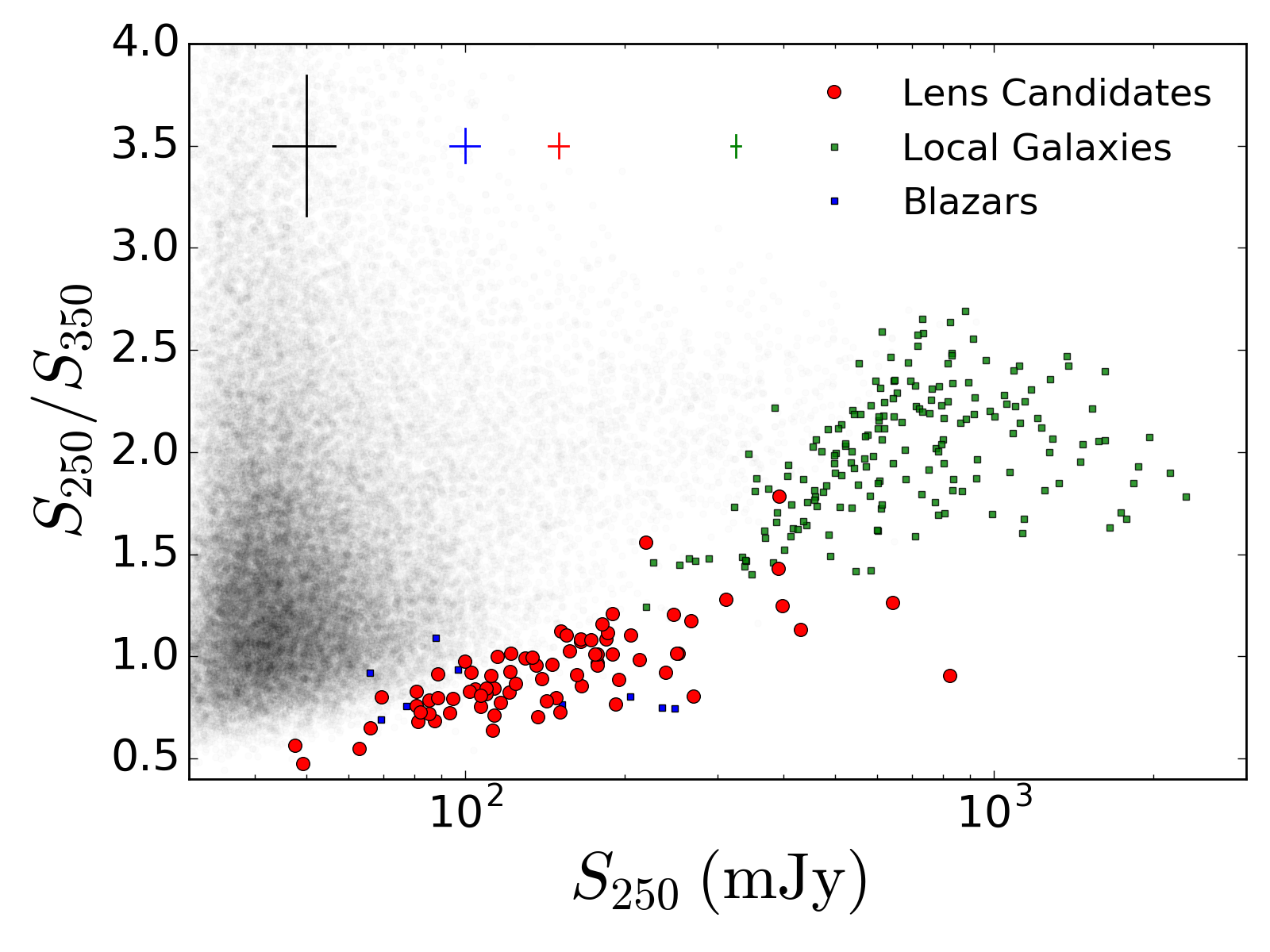}
\includegraphics[scale=0.4]{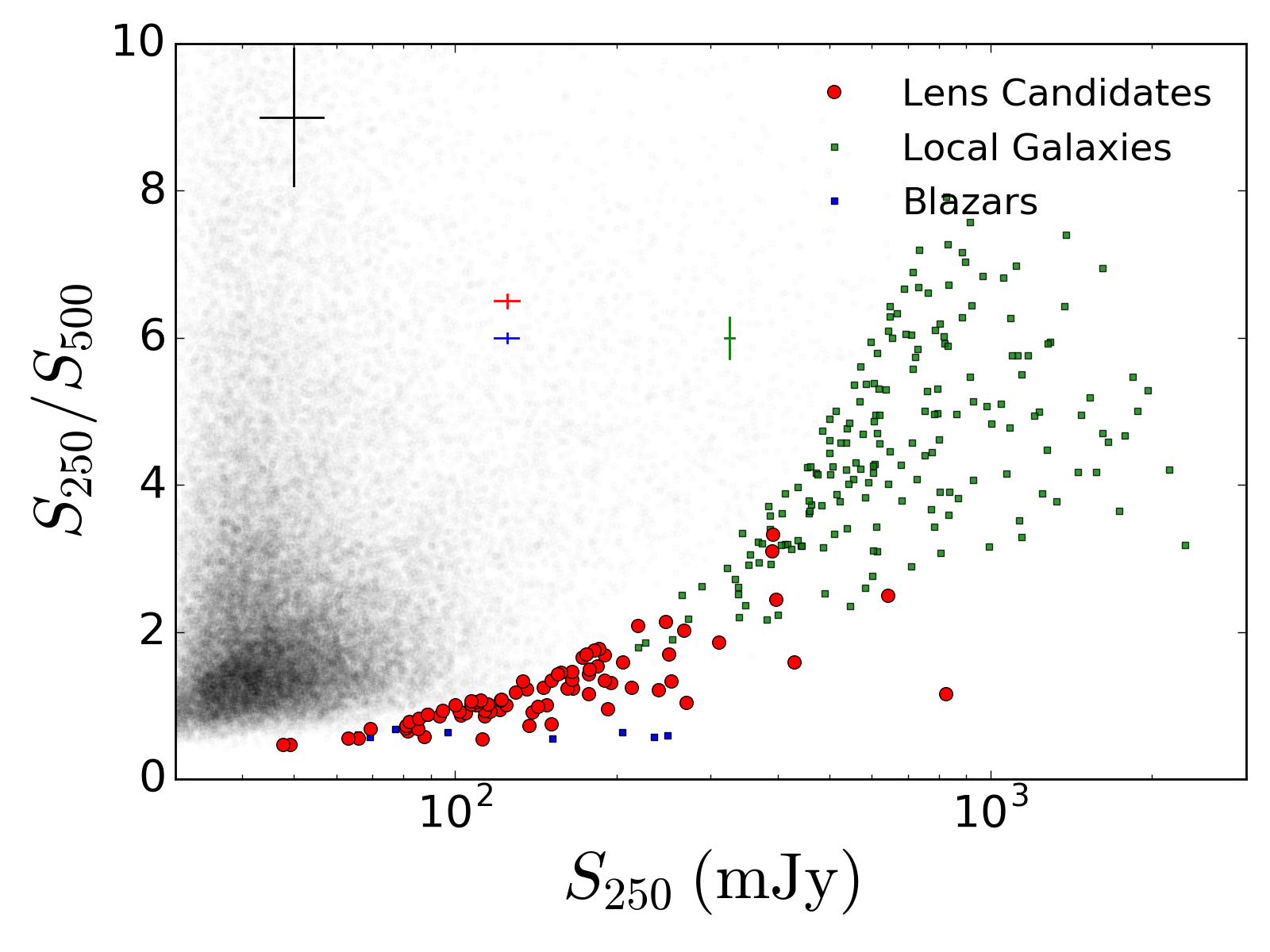}
\caption{{\it Herschel}/SPIRE color vs. flux density plots for all sources in the HeLMS and HerS catalogs. Lens candidates, local galaxies, and blazars with $S_{500}>$\,100\,mJy are shown in color while all sources are shown in gray scale. Median error bars for all four populations are given at the top of the figures. The DSFGs in our gravitationally lensed candidate catalog are redder than the local galaxies, consistent with a greater redshift. The apparent offset of the lens candidates from the gray scale general population is due to our flux cut at $S_{500}=100$\,mJy and vanishes in the color-color plot (Figure 4).}
\end{figure}

\begin{figure}
\centering
\includegraphics[scale=0.4]{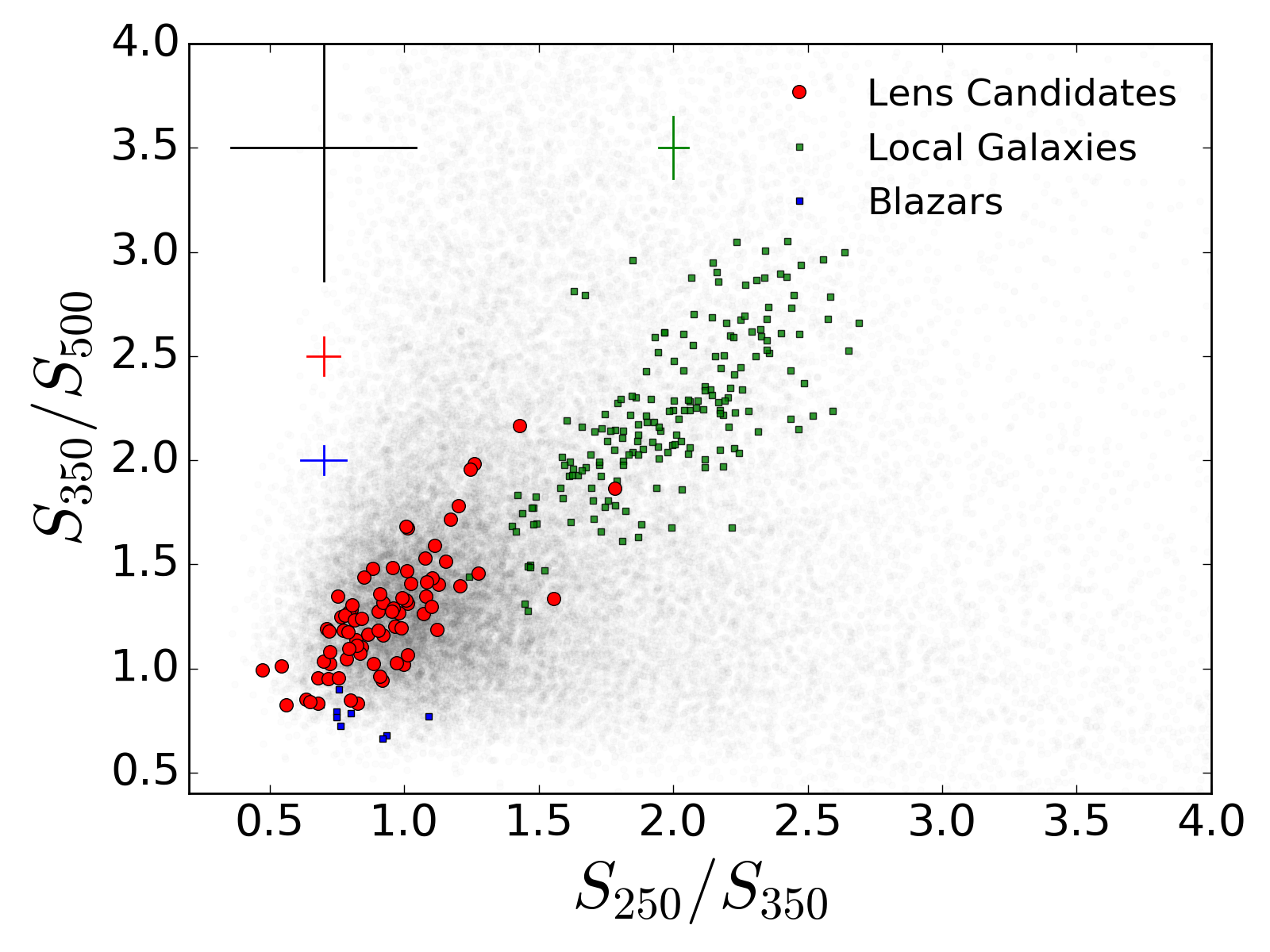}
\caption{Color-color plot for all sources in the HeLMS and HerS catalogs. Sources with $S_{500}>100$\,mJy are highlighted when they belong to our lensing candidate, local galaxy, or blazar catalogs. The background gray scale shows the distribution of all sources in HeLMS and HerS. The colors of our candidate lensed DSFGs match those of the general HeLMS and HerS population. The error bars are standard deviations of the different populations.}
\end{figure}

\begin{figure}
\centering
\includegraphics[width=8.5cm]{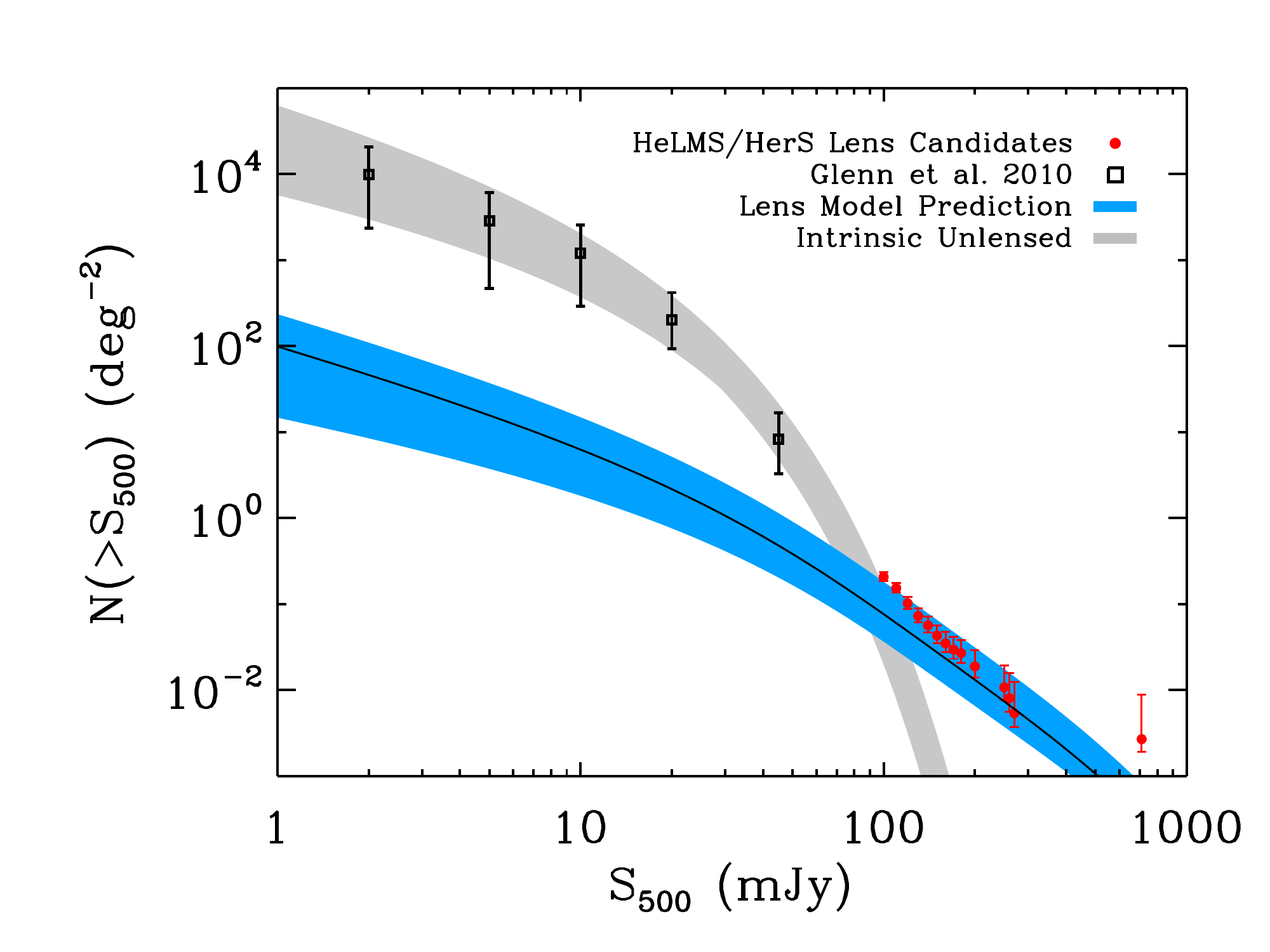}
\caption{Cumulative 500\,$\mu$m number counts as a function of the {\it Herschel}/SPIRE 500\,$\mu$m flux. The lensed galaxy candidate counts from HerS and HeLMS are overlaid in red. These are consistent with counts of lensed sources as predicted from lensing statistics \citep{Wardlow2013} shown with the black line with confidence interval in blue. For comparison we show the unlensed source counts from the HerMES blank-field catalogs \citep{Oliver2012} and from P(D) analysis \citep{Glenn2010} as grey shaded area and black squares respectively. The total counts of {\it Herschel} detected sources in these fields will be presented in future studies (Oliver et al., in prep, Clarke et al., in prep).}
\end{figure}

\begin{figure*}
\centering
\includegraphics[trim=0cm 0cm 0cm 0cm, scale=0.7]{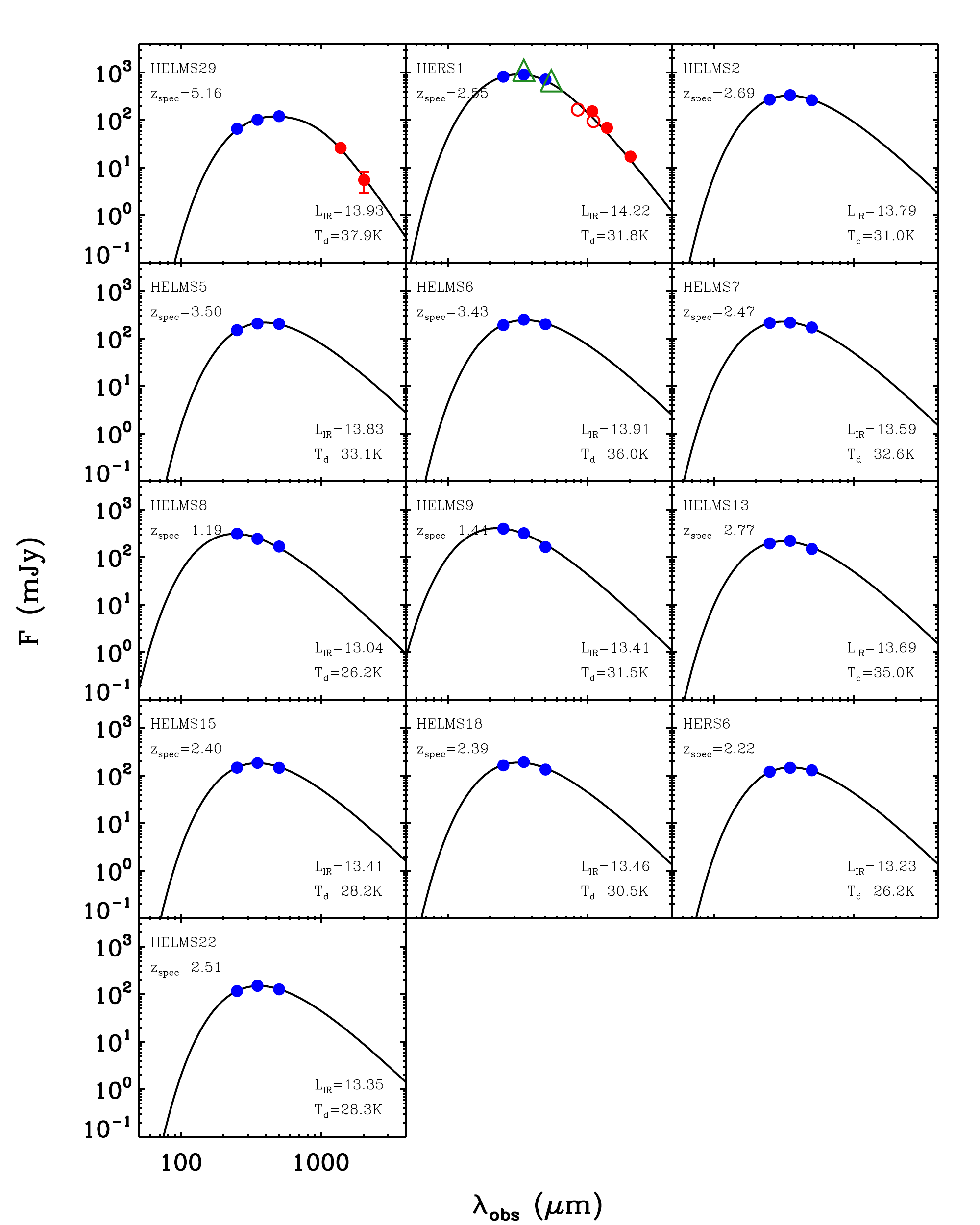}
\caption{Far-infrared SED fits of the thirteen HerS and HeLMS lens candidates with CO spectroscopic redshifts using a modified black-body fit with $\beta=1.5$ \citep{Casey2012a}. First two sources show mm-band observations from the Atacama Cosmology Telescope \citep{Su2015} in filled red circles. HERS1 also has observations in mm-bands from \citet{Geach2015} (open circles) and also data from Planck point source catalog that is represented as open green triangles. The total infrared luminosity (integrated over 8-1000 $\mu$m in units of ${\rm Log(L_{\odot})}$) and far infrared measured dust temperatures from the black-body fits (${\rm T_d}$) are also reported for each object.}
\end{figure*}

\begin{figure}
\centering
\includegraphics[trim=2.5cm 0cm 0cm 0cm, scale=0.45]{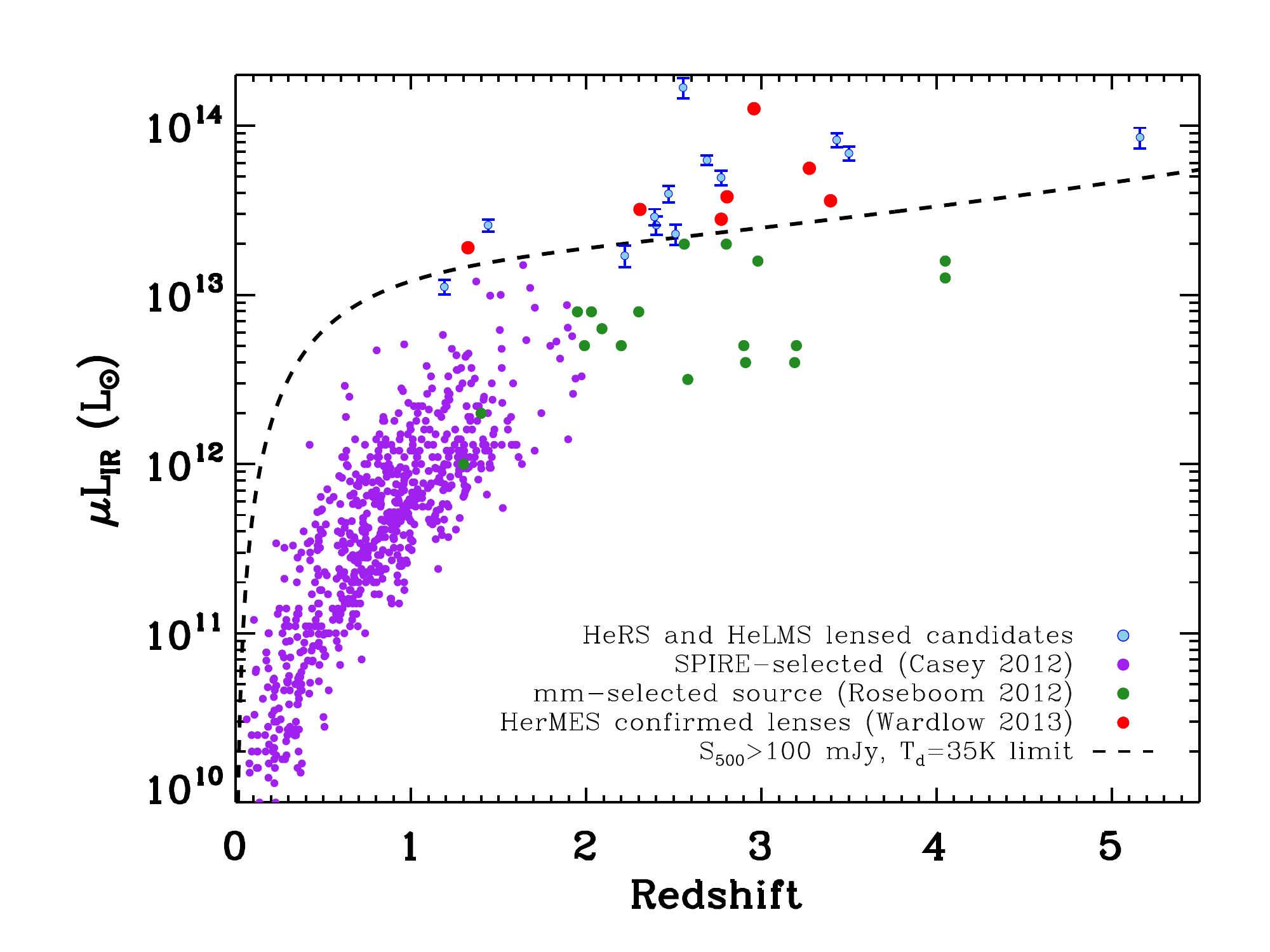}
\caption{Total observed infrared luminosity ($\rm \mu L_{IR}$; rest-frame $\rm 8-1000\,\mu m$) measured from the {\it Herschel}/SPIRE data as a function of redshift. The DSFG redshifts are from spectroscopic CO observations by CARMA (Reichers et al., in prep) and the GBT (Harris et al., in prep). The curved dashed line shows the detection limit for our candidate lensed DSFGs from a modified blackbody model with $\beta = 1.5$, $\rm T_d = 35\,K$ and 500\,$\rm \mu m$ flux of 100\,mJy similar to our selection flux cut (see \citealp{Casey2012a}). For comparison we are showing SPIRE-selected and mm-selected samples of star forming galaxies from \citet{Casey2012b} and \citet{Roseboom2012} respectively. Candidate lensed DSFGs identified by \citet{Wardlow2013} are shown with filled red circles. At any given redshift, the lensed DSFGs would have a total infrared luminosity that exceeds those of red star forming galaxies because of the lensing magnification.}
\end{figure}

\begin{figure}
\centering
\includegraphics[trim=2.5cm 0cm 0cm 0cm, scale=0.45]{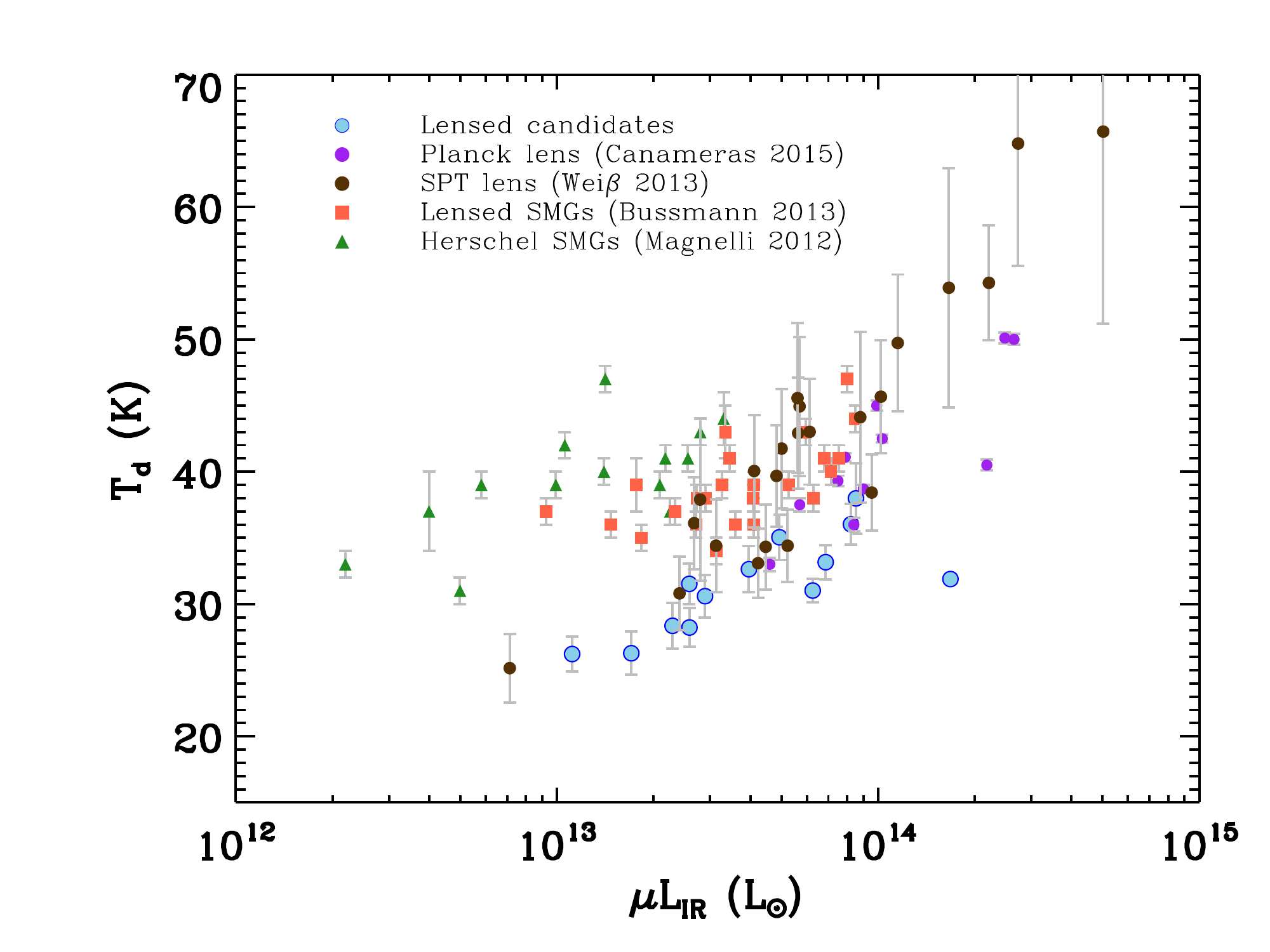}
\caption{Dust temperature vs. total infrared luminosity for the candidate gravitationally lensed DSFGs in the HeLMS and HerS. The temperature and $\rm \mu L_{IR}$ (rest-frame $\rm 8-1000\,\mu$m) are plotted for a sub-sample of the lens candidates for which we have spectroscopic redshifts from CO observations by fitting a modified blackbody spectrum (see \citealp{Casey2012a}). The dust temperatures and infrared luminosities of several studies of lensed DSFGs and SMGs are shown for comparison \citep{Magnelli2012, Bussmann2013, Weiss2013, Canameras2015}. The smaller scatter in the derived dust temperatures is associated with the use of spectroscopic redshift information, from CO observations, in the SED fits. The lensed DSFGs have median dust temperatures of $\rm T_d\sim 31\,K$ indicating a secular mode of star formation.}
\end{figure}

\begin{longtable}{*{4}{c}}
\caption{Cumulative number density of lensed sources.}\\
\hline
Flux Limit & Number Density & Low 68\%$^{\dagger}$ & High 68\%$^{\dagger}$ \\
500\,$\mu$m (mJy) & (deg$^{-2}$) & (deg$^{-2}$) & (deg$^{-2}$) \\
\hline
\hline
100  &   0.207  &   0.186   &  0.233 \\
110   &  0.153   &  0.136   &  0.176 \\
120  &   0.102  &  0.088   &  0.121 \\
130  &   0.073  &  0.061  &  0.089 \\
140  &   0.056  &  0.047  &  0.072 \\
150  &   0.043  &  0.035  &  0.057 \\
160  &   0.035  &  0.028  &  0.047 \\
170   &  0.030  &  0.023  &  0.041 \\
180   &  0.027  &  0.021  &  0.038 \\
200   &  0.019  &  0.014  &  0.029 \\
250   &  0.011 &  0.008  &  0.019 \\
260   &  0.008 &  0.006 &   0.016 \\
270   &  0.005 &  0.004  &  0.012 \\
710   &  0.0027 &  0.0019 &  0.0088 \\
\hline
\end{longtable}
{\footnotesize $^{\dagger}$: Confidence intervals are calculated following the prescription of \citet{Gehrels1986}.}

\section{Follow-up Observations}

\subsection{Redshifts}

The redshifts of gas-rich DSFGs can be determined using CO rotational emission lines in radio observations. This method has been used extensively to measure the redshift of distant DSFGs \citep{Greve2005, Swinbank2010b, Lupu2012, Harris2012, Weiss2013, George2013, Decarli2014, Canameras2015, Zavala2015}. The CO molecular line observations are also used to study the physical propertied of SMGs and DSFGs \citep{Tacconi2006, Tacconi2008, Solomon2005, Carilli2013}. In particular CO brightness and spatial distribution could be utilized to determine the total molecular gas content and extent, which is responsible for star formation, in these gas-rich systems \citep{Narayanan2012, Fu2012, Fu2013}. As part of existing programs to follow-up {\it Herschel}-selected bright lensed galaxies, we obtained radio/millimeter CO redshift measurements of 12 sources in our HeLMS+HerS sample with the Robert C. Byrd Green Bank Telescope (GBT; 7 sources; Harris et al., in prep.), the Combined Array for Research in Millimeter-wave Astronomy (CARMA; 12 sources; Riechers et al., in prep.), and the IRAM/Plateau de Bure Interferometer (PdBI; 11 sources; Riechers et al., in prep), and one source (HELMS29) with spectroscopic observations from ALMA \citep{Asboth2016}. The 12 sources with redshifts reported here come from a large sample of lensed {\it Herschel} sources, involving all of the wide area {\it Herschel} fields. The CO sample follows no particular selection criterion apart from flexibility if scheduling with each of the facilities. The GBT observations target CO\,($1 \rightarrow 0$) transition, however as discussed in \citet {Harris2012}, it is also possible to have a GBT CO\,($2 \rightarrow 1$) transition for systems at $5.1<z<8$ or a line from species other than the CO, however this would be unlikely given the photometric redshift distributions and the weak nature of the other line species \citep{Harris2012}. The lensed DSFG candidates with confirmed spectroscopic redshift from CO observations have 500\,$\mu$m fluxes exceeding 121\,mJy with a median flux of 164\,mJy in the 500\,$\mu$m. The spectroscopic observations confirmed the brighter candidates in the far-infrared demonstrating increased chance of detection for brighter objects (Riechers et al., in prep, Harris et al., in prep). 

Figure 1 shows the redshift distribution of the background DSFGs measured from CO. The measured redshifts of the background sources are at $z>1$ and peak at $z \sim 2.5$. This result is consistent with other studies of SMG redshift distributions \citep{Chapman2005, Harris2012, Bussmann2013, Bussmann2015}. Figure 1 also shows the redshift distribution of the foreground lensing galaxies. This result is derived from public SDSS catalogs and is mostly for the brighter optical counterparts (due to limited spectroscopic depth). However our spectroscopic campaigns on the {\it Herschel} selected lensing systems have successfully measured the individual galaxy spectra using large ground-based observatories such as Keck (DEIMOS and LRIS) along with observations from Gemini, MMT and VLT \citep{Bussmann2013}. The redshift distribution of the foreground galaxies peaks at $z<1$ and is consistent with the scenario that a background high redshift galaxy is being gravitationally lensed by a foreground system at much lower redshift. 

\subsection{Imaging}

As part of a ``HELMS Deep'' observing campaign, several of the lens candidates have been followed up with high resolution imaging with Keck/NIRC2 laser-guided adaptive optics (LGSAO) and with seeing limited imaging with the WHT LIRIS instrument in the near infrared. These observations were designed to study the rest-frame optical properties of these lensed systems in detail, as has been done over the past few years for other DSFGs at $z>1$ \citep{Fu2012, Fu2013, Calanog2014, Timmons2015}. HELMS Deep observations
of the lensed candidates were obtained with the Keck/NIRC2 AO system in the $K_s$ band at 2.2\,$\mu$m in August and September of 2015 (PI: Cooray). The observations were done with an average exposure time of 3600 sec over two nights under cloudy conditions and poor AO correction. 
Figure 2 shows the KeckII/NIRC2-LGSAO K$_s$-band images of two of the lensed systems for 
which we obtained reliable data in these runs that are based on the new catalog presented here. 
The two galaxies (HERS1, discussed below and identified elsewhere as a lensing galaxy, and HERS6) both show clear lensing features in the near-infrared composed of arcs and counter images. A detailed analysis of these imaging data, in combination with other multi-wavelength interferometric images, including
those from an ALMA snapshot program (PI: Eales) will be presented in future papers. We highlight them here to motivate additional follow-up programs of the {\it Herschel} lens sample by the strong lensing community.

The WHT/LIRIS observations of HeLMS lens candidates are part of a Large Program on 
HerMES high-redshift galaxies (PI: Pérez-Fournon). The LIRIS observations were 
done with a typical exposure time of 3600 sec in sub-arcsecond seeing conditions. Figure 2 also
shows the WHT/LIRIS $K_s$ imaging of HELMS6 observed on 26 October 2015, with seeing of 0.8$''$. 
In this case, the main lensing galaxy is a member of a cluster of galaxies, with SDSS photometric 
redshift $\sim 0.4$. Giant arcs are also present in this system. Optical long-slit spectroscopy 
of this field with the OSIRIS instrument of the Gran Telescopio Canarias (GTC) was obtained 
on 29 November 2015 in dark and clear conditions and seeing of 1.2 arcseconds. The exposure time 
was 3000 sec. We used the R1000B grism, with a spectral sampling of 2.12\,$\rm \AA\,pixel^{-1}$, and a slit 
width of 1.2 arcseconds. The slit included several of the galaxies in the cluster to the south of the 
main lensing galaxy. This position was chosen to include one of the lensed arc features. 
From these observations we could measure the spectroscopic redshifts of three galaxies in the cluster, 
of 0.3950, 0.3956, and 0.3968, close to the SDSS photometric redshift. More details and analysis 
will be presented in Marques-Chaves et al. (in prep).

\section{Discussion}

\subsection{Far-Infrared Colors}

Figure 3 shows 250, 350, and 500\,$\mu$m color-flux density plots for all sources detected in the HeLMS and HerS catalogs. Sources of interest, with $S_{500}>100$\,mJy, are highlighted in red, blue, and green, for our lens candidate galaxies, local galaxies, and blazars, respectively. As mentioned in Section 2, all flux measurements are taken from the HeLMS and HerS catalogs referenced above. 

Our primary catalog of lensed DSFG candidate galaxies has median $S_{350}/S_{500}=1.29 \pm 0.10$, $S_{250}/S_{350}=0.90 \pm 0.06$, and $S_{250}/S_{500}=1.16 \pm 0.09$ with errors being the standard deviation of the distribution. In comparison, the general population of all sources has median $S_{350}/S_{500}=1.54 \pm 0.63$, $S_{250}/S_{350}=1.34 \pm 0.33$, and $S_{250}/S_{500}=2.06 \pm 0.80$. These have comparable colors, as expected, since the lensed candidates are drawn from the parent population of {\it Herschel} detected sources. The local galaxies have bluer colors than either the general population or our candidate lensed DSFGs, with median  $S_{350}/S_{500}=2.29 \pm 0.13$, $S_{250}/S_{350}=2.07 \pm 0.08$, and $S_{250}/S_{500}=4.74 \pm 0.27$. As this population is exclusively composed of galaxies at very low redshift, the relatively blue color is expected. The blazars have median $S_{350}/S_{500}=0.68 \pm 0.08$, $S_{250}/S_{350}=0.93 \pm 0.08$, and $S_{250}/S_{500}=0.63 \pm 0.05$.

Our highlighted lens candidates, local galaxies, and blazars are offset from each other in the $S_{250}/S_{350}$ vs $S_{250}$, $S_{250}/S_{500}$ vs $S_{250}$, and $S_{350}/S_{500}$ vs $S_{500}$ color-flux space due to our selection criteria. Since we limited our search to sources bright in the $S_{500}$, which are correspondingly bright in the $S_{250}$, the highlighted sources all appear shifted to the right in Figure 3. As we see in Figure 4, this effect is not present in the color-color space where the distribution of candidate galaxies follows the general population. 

\subsection{Lensing Statistics}

The statistics of lensing at sub-millimeter wavelengths has been discussed extensively in \citet {Perrotta2002, Negrello2007, Lima2010, Hezaveh2011, Wardlow2013}, where similar techniques were used to identify lensed DSFGs in HerMES using {\it Herschel} 500\,$\mu$m observations. The adopted model assumes a foreground mass profile and spatial distribution for the lensing galaxies, along with a redshift distribution of DSFGs with $S_{500}>1$\,mJy, adopted from \citet{Bethermin2012b} for un-lensed DSFGs. The model ultimately makes predictions regarding the properties of lensed DSFGs \citep{Wardlow2013}. Figure 5 shows the cumulative 500\,$\mu$m number counts of galaxies, considering the assumptions above. The cumulative number counts of lensed candidates are reported in Table 1. The confidence intervals in these measurements are calculated following the prescription of \citet{Gehrels1986}. We see a steep slope in the 500\,$\mu$m number counts of un-lensed DSFGs, showing that the population of bright 500\,$\mu$m sources should be dominated by gravitationally lensed objects along with bright local spirals in the 500\,$\mu$m data \citep{Wardlow2013}. 

According to the same lensing model, a 100\,mJy flux cut produces a strong-lensing galaxy fraction of $32\%-74\%$ \citep {Wardlow2013} with an intrinsic flux density distribution that peaks at 5\,mJy for the DSFGs. Given the HeLMS and HerS detection limits, many of these sources would be undetected without the 
magnification boost coming from lensing. Another prediction of this statistical model is that source blending does not significantly affect the lensed DSFG selection. This takes into account the blending of several intrinsically faint sources within the {\it Herschel} beam size that could produce a source with a combined flux above 100\,mJy at 500\,$\mu$m. Based on simulations, \citet{Wardlow2013} estimated that the blending of fainter galaxies has a probability of less than $5 \times 10^{-5}$ to account for a source with flux density greater than 100\,mJy. Within our sample of 77 lensed candidates, we do not expect a single source to result from the blending of two random fainter galaxies and still be identified as a single source. This is a result of the rarity of the bright DSFGs. 

This estimate, however, ignores physical associations, rare instances of DSFG-DSFG mergers, or especially SMG-SMG mergers similar to the ones already uncovered with {\it Herschel} (eg. \citealp{Fu2013, Ivison2013}). Such sources are now believed to make up a considerable fraction of the $S_{500} >100$\,mJy population. Among the 13 lensed candidates in HerMES \citep{Wardlow2013}, one source was identified to be an SMG-SMG merger at $z=2.3$ using the Sub-Millimeter Array (SMA) and other follow-up observations (HXMM01 of \citealp{Fu2013}). Using ALMA in Cycle 0 (see \citealp{Bussmann2015}) an additional source from the group of 13 was found to be a blend of  at least three DSFGs that are weakly to moderately lensed (with $1<\mu<2$) by a low redshift galaxy that is 6 arcseconds away from the centroid of the three ALMA-detected sources.
While we do not have precise predictions, we expect $10\%-15\%$ of the sample to be SMG-SMG mergers. Detailed follow-up studies of the statistically significant sample presented here, involving a total of 77 lensed candidates, can be used to estimate the fraction more precisely. The exact fraction of SMG-SMG mergers, and their luminosity and mass distribution, is crucial for connecting such mergers with the formation history of the most massive and red galaxies at $z > 2$.

\subsection{Luminosities and SEDs}

For the subsample of lensed candidate DSFGs from HeLMS and HerS with CO-based redshifts, we deduce the observed total infrared luminosity (${\rm L_{\rm IR}}$; restframe 8-1000 $\mu$m) from the {\it Herschel}/SPIRE observations at 250, 350 and 500\,$\mu$m using modified black-body fit to the data as explained in \citet{Casey2012a}. The modified black-body takes into account variation in opacity and source emissivity. These fits are summarized in Figure 6. Two of the sources overlap with \citet{Su2015}, as discussed below, and we include their flux densities. For one of the targets, HERS1, we also include Planck-detected flux densities.

Figure 7 shows ${\mu \rm L_{\rm IR}}$ as a function of spectroscopic redshift measured from CO observations. The DSFGs have measured total infrared luminosities between $1.1 \times 10^{13}\,\mu^{-1}{\rm L_{\odot}}$ and $1.6 \times 10^{14}\,\mu^{-1}{\rm L_{\odot}}$ and a median of $4.9 \times 10^{13}\,\mu^{-1}{\rm L_{\odot}}$. Using a \citet{Kennicutt1998} relation for star formation (${\rm SFR (M_{\odot}yr^{-1})=4.5\times10^{-44}\,L_{IR}\,(ergs\:s^{-1})}$ with a Salpeter initial mass function; \citealp{Salpeter1955}) we infer a median star formation rate of $\sim 8500\,\mu^{-1}\text{M}_{\odot}\text{yr}^{-1}$. In the above $\mu$ is the gravitational lensing magnification for the {\it Herschel} selected lensed DSFGs. We can estimate the average maximum magnification as a function of the 500\,$\mu$m flux from statistical gravitational lens modeling \citep{Lapi2012, Wardlow2013} and direct observations of large SMA-detected samples of lensed DSFGs and SMGs \citep{Bussmann2013}. The measured infrared luminosities for these targets far exceed those of normal star-forming and IR luminous systems, providing further evidence of gravitational lensing, which is consistent with previous studies of lensed DSFGs \citep{Wardlow2013}. In fact, given the average magnification for bright gravitationally lensed sources \citep{Wardlow2013, Bussmann2013} we infer a total infrared luminosity of lensed DSFGs consistent with that of normal IR-bright galaxies at similar redshifts.  

Figure 8 shows the dust temperature, derived from SED fits described above, as a function of the total infrared luminosity for our lensed DSFG candidates with spectroscopic redshifts in the HeLMS and HerS fields. The candidate lensed systems have observed dust temperatures $\sim 25-40$\,K and total observed luminosities (${\mu \rm L_{\rm IR}}$) as reported above. Figure 8 also shows the dust temperature and infrared luminosities of other samples of lensed DSFGs and SMGs from the literature \citep{Magnelli2012, Weiss2013, Bussmann2013, Canameras2015} where we have converted the intrinsic values to observed properties given the derived magnifications in those studies. The spread in the distribution of these sources on the $\rm T_d$ vs $\rm \mu L_{IR}$ plane is mostly associated with the different selection functions and also intrinsic properties of each population. Our candidate DSFGs show a smaller scatter in the measured properties which is mainly associated with the use of the redshift information (from the CO observations) for these systems in our SED fits \citep{Magnelli2012}. Our DSFG candidates have median SED measured dust temperatures of $\rm T_d \sim 31$\,K. This indicates that our candidate DSFGs are more dominated by systems with cooler dust temperatures which is more consistent with secular modes of star formation rather than merger driven star formation activity \citep{Magnelli2012}, also evident from our high resolution follow-up observations. 

\subsection{Example Lensed Sources}

HERS1, shown in Figure 2, had been already identified as a lensed galaxy by a citizen science group {\sc SpaceWarps} making use of VISTA-CFHT optical imaging data in SDSS Stripe 82 \citep{Geach2015}. It is the brightest target in our list of 77 with a 500\,$\mu$m flux density of 717\,mJy. 
It is a unique source in that it stands out at the bright-end separately on its
own while the next set of lensed candidate 
sources have flux densities well below 300\,mJy. The Keck/NIRC2 LGS-AO HELMS Deep K$_s$-band we present in Figure 2, obtained prior to us knowing about the {\sc SpaceWarps} project and the \citet{Geach2015} publication, has substantially better resolution than VISTA-CFHT $K_s$-band image in \citet{Geach2015}. 
Unlike our previous lens follow-up programs in the $K$-band that resulted in the need for deep integrations due to faintness of the lensed source, HERS1 is bright enough in the $K$-band that the lensing galaxy is easily visible in a single NIRC2 frame of 60 seconds.

Additional multi-wavelength observations of this target, including CO spectroscopy, is reported in \citet{Geach2015}. HERS1 is also detected in Planck all-sky point source catalog and is identified as PCCS2 857 G160.58-56.76 in the Planck 857\,GHz point source catalog with a separation of less than 1 arcminute from the {\it Herschel}/SPIRE 250\,$\rm \mu m$ position, which is smaller than full-width-half-maximum (FWHM) resolution of the Planck ($\sim\,4^{\prime}$) at 857\,GHz. The Planck has a 857\,GHz flux of 1118 $\pm$ 408\,mJy and a 545\,GHz flux of 676 $\pm$ 197\,mJy. These flux densities are consistent with the SPIRE 350\,$\rm \mu m$ and 500\,$\rm \mu m$ flux of 912 $\pm$ 7\,mJy and 718 $\pm$ 8\,mJy, respectively. The SED is shown in Figure 6. Fixing $\beta = 1.5$ does not provide a good fit for HERS1 given the additional data that we have, in particular at longer wavelengths. We kept $\beta$ as a free parameter \citep{Casey2012a} while fitting the full SED of HERS1 with {\it Herschel}/SPIRE, Planck and ACT data, giving a $\beta=1.9$ and providing a better fit.

In a recent study \citet{Su2015} identified a sample of nine gravitationally lensed DSFGs using the Atacama Cosmology Telescope (ACT) in Stripe 82. Six of the nine lensing galaxies have $S_{500}>100$\,mJy. We had independently selected these six candidates in our sample, prior to our knowledge of the \citet{Su2015} study or preprint, in the current sample of HerS/HeLMS candidate gravitationally lensed galaxies. The three targets not selected in our study have $S_{500}<100$\,mJy and fall outside our lensed DSFG candidate selection. The far-infrared fluxes of all the 12 ACT candidates are consistent with the {\it Herschel}/SPIRE fluxes of the corresponding sources in our catalog, given the errors and the leeway in the SED fits. Two of the ACT detected lensed galaxies in our catalog (HERS1 and HELMS29) have spectroscopic redshift from CO observations (Harris et al in prep, Riechers et al in prep). HELMS29 at $z=5.162$ (see below) is the highest redshift lensing
galaxy in our sample. We use this redshift along with the {\it Herschel} far-infrared and ACT mm data to construct the SED (Figure 6). Similar to HERS1, discussed above, we treated $\beta$ as a free parameter in the SED fit of HELMS29. The addition of ACT data provided a better fit to the flux densities at longer wavelengths with a $\beta=2.6$ (for more details on the fitting procedure see: \citealp{Casey2012a}). \citet{Su2015} also presents deep follow-up observations of their lensing sample, including CO redshift of one additional source (their other redshift is for HERS1). Unfortunately that source does not fall within the $S_{500}>100$\,mJy limit we have used in the present catalog. These independent follow-up campaigns by different groups (which we label as ``battle of HELMS deep'') will likely result in additional redshifts for the lens galaxy sample we have presented here.

\citet{Asboth2016} discuss the ``red'' galaxies in HeLMS with $S_{500} > S_{350} > S_{250}$. Such a selection identifies candidate SMGs with $z>4$, similar to the selection that led to HFLS3 \citep{Riechers2013, Cooray2014}, as discussed in \citet{Dowell2014}. Six targets overlap between the two samples and are identified here in Table 2. These sources are likely to be lensed galaxies with $z > 4$. One such candidate, HELMS29, has a CO-based redshift of 5.162 and is discussed in detail in \citet{Asboth2016}.
The redshift is also confirmed by an ALMA Cycle 1 observation (PI: Conley), as discussed in \citet{Asboth2016}. The HeLMS and HerS candidate list presented here is also part of a ALMA Cycle 2 snapshot program (PI: Eales). Ten of the targets from the HeLMS list have secured observations at 350\,GHz and those targets are identified in Table 2. The ALMA data, lensing models, and multi-wavelength analysis will be presented in future papers.

\section{Summary and Conclusions}

We used {\it Herschel}/SPIRE maps of HeLMS and HerS surveys to produce a list of candidate gravitationally lensed DSFGs. Our main findings are:

\begin{itemize}

\item We identified 77 candidate lensed dusty star forming galaxies in the combined 372\,deg$^{2}$ region covered by the {\it Herschel} HeLMS and HerS fields (0.21 $\pm$ 0.03 $\rm deg^{-2}$).

\item Candidate DSFGs are bright in the SPIRE 500\,$\mu m$ observations ($S_{500}>100$\,mJy by selection). We further show that the colors of our candidate DSFGs are redder than the local spiral galaxies, indicative of a separate population at high redshift.

\item Few of our brighter lensed DSFGs with red SEDs in the HeLMS field have been recently discovered in a parallel study \citep{Asboth2016}. The Atacama Cosmology Telescope (ACT) also confirmed the existence of a small sub-sample of our lensed DSFG candidates \citep{Su2015}. This further demonstrates the robustness of our selection of high redshift lensed galaxies. 

\item High resolution near-infrared follow-up observations of a few of the candidates with Keck/NIRC2 AO and William Herschel Telescope (WHT) reveal arcs and distorted images associated with gravitational lensing.

\item Spectroscopic redshifts measured from CO molecular line observations (Harris et al., in prep; Riechers et al., in prep) for thirteen of the candidates put the population of lensed DSFGs at $z>1$ whereas the foreground lensing galaxies have redshift distribution that peaks at $z<1$. This further supports the scenario of a population of distant objects being gravitationally lensed by nearby foreground galaxies. 

\item We fit the far-infrared SED of the candidate lensed DSFGs with modified black-body which gives us the best-fit total infrared luminosity ($\rm L_{IR}$; rest-frame $\rm 8-1000\,\mu m$) and dust temperature ($\rm T_d$). For the SED fittings we fix the redshift of the galaxy to spectroscopic redshifts from CO observations. Two of the candidate DSFGs (HERS1 and HELMS29) also have longer wavelength data from ACT observations with HERS1 being also detected in the Planck point source catalog. We use these additional data in our SED analysis.

\item The lensed DSFGs have median total infrared luminosities of $4.9 \times 10^{13}\:\mu^{-1}{\rm L_{\odot}}$ where $\mu$ is the gravitational lensing magnification factor. The observed infrared luminosity for lensed DSFGs far exceeds that of normal star forming galaxies at similar redshifts. In fact given the average magnification for bright gravitationally lensed sources \citep{Bussmann2013} we infer a total infrared luminosity of lensed DSFGs consistent with that of normal IR-bright galaxies at similar redshifts.

\item The candidate lensed galaxies have SED measured dust temperatures in the range of $\rm 25\,K<T_d<38\,K$ with a median of $\rm T_d\sim 31\,K$ and a small scatter given the spectroscopic CO information. The relatively low dust temperature compared to the merger-driven dust temperatures in the IR luminous galaxies at similar redshifts supports the secular star formation evolution within these systems.

\end{itemize}

Given the high success rate of high redshift DSFG observations selected similarly from other studies \citet{Negrello2010, Wardlow2013} and from our own follow-up high resolution observations, we can conclude that the present catalog is a robust list of bona-fide gravitationally lensed galaxies at $z>1$. About 10\% to 15\% of the sample is likely luminous SMG-SMG mergers that are of interest for understanding the formation paths of massive galaxies at $z > 2$. The lensing nature, properties of the lensed galaxies, and identification of SMG-SMG mergers for further studies will require carefully planned follow-up campaigns with a variety of facilities in the future.

\section*{Acknowledgement} 

We wish to thank the anonymous referee for thoroughly reading the manuscript and providing very useful suggestions. Financial support for this work was provided by NSF
through AST-1313319 for HN and AC and NSF REU support in AST-1313319 for MK. UCI group also acknowledges NASA support for Herschel/SPIRE GTO and Open-Time Programs.
RJI acknowledges support from the European Research Council in the form of the Advanced Investigator Programme, 321302, COSMICISM. JLW is supported by a European Union COFUND/Durham Junior Research Fellowship under EU grant agreement number 267209, and acknowledges additional support from STFC (ST/L00075X/1). Support for IPF, PIMN and RMS was from the Spanish MINECO grants AYA2010-21697-C05-4, 
FIS2012-39162-C06-02, and ESP2013-47809-C3-3-R. SO acknowledges support from the
Science and Technology Facilities Council (grant number
ST/L000652/1). The research leading to these results has received funding from the European Union Seventh Framework Programme FP7/2007-2013/ under grant agreement n$^\circ$ 607254. Disclaimer: ``This publication reflects only the author’s view and the European Union is not responsible for any use that may be made of the information contained therein.'' Some of the data presented herein were obtained at the W.M. Keck Observatory, which is operated as a scientific partnership among the California Institute of Technology, the University of California and the National Aeronautics and Space Administration. The Observatory was made possible by the generous financial support of the W.M. Keck Foundation. The authors wish to recognize and acknowledge the very significant cultural role and reverence that the summit of Mauna Kea has always had within the indigenous Hawaiian community. We are most fortunate to have the opportunity to conduct observations from this mountain. Data presented herein were obtained using the UCI Remote Observing Facility, made possible by a generous gift from John and Ruth Ann Evans. The {\it Herschel} spacecraft was designed, built, tested, and launched under a contract to ESA managed by the {\it Herschel}/Planck Project team by an industrial consortium under the overall responsibility of the prime contractor Thales Alenia Space (Cannes), and including Astrium (Friedrichshafen) responsible for the payload module and for system testing at spacecraft level, Thales Alenia Space (Turin) responsible for the service module, and Astrium (Toulouse) responsible for the telescope, with in excess of a hundred subcontractors. HCSS / HSpot / HIPE is a joint development (are joint developments) by the {\it Herschel} Science Ground Segment Consortium, consisting of ESA, the NASA {\it Herschel} Science Center, and the HIFI, PACS and SPIRE consortia. The IRAM Plateau de Bure
Interferometer is supported by INSU/CNRS (France), MPG
(Germany), and IGN (Spain). The National Radio Astronomy
Observatory is a facility of the National Science Foundation
operated by Associated Universities, Inc. Support for CARMA
construction was derived from the Gordon and Betty Moore
Foundation, the Kenneth T. and Eileen L. Norris Foundation,
the James S. McDonnell Foundation, the Associates of the California
Institute of Technology, the University of Chicago, the
states of California, Illinois, and Maryland, and the National
Science Foundation. Ongoing CARMA development and operations
are supported by NSF grant ATI-0838178 to CARMA,
and by the CARMA partner universities. The William Herschel telescope is operated on the island of La Palma by the Isaac Newton Group 
in the Spanish Observatorio del Roque de los Muchachos of the Instituto de Astrofisica de Canarias. 
Based in part on observations made with the Gran Telescopio Canarias (GTC), installed
at the Spanish Observatorio del Roque de los Muchachos of the Instituto de Astrofisica de Canarias, 
in the island of La Palma.

\bibliographystyle{apj}
\bibliography{draft}

\newpage

\section*{Appendix}




\begin{longtable*}{llllll}
\caption{HeLMS lens candidates ($S_{500}>100$\,mJy)}\\
\hline

Object ID & RA & Dec & $S_{250}$\,(mJy) & $S_{350}$\,(mJy) &
$S_{500}$\,(mJy) \\
\\
\hline

HELMS1 & $23^{\rm h}34^{\rm m}41^{\rm s}.0$& $-06\degree52'20''$&  431 $\pm$ 6  &  381 $\pm$ 7  &  272 $\pm$ 7   \\

HELMS2$^{\ddagger}$ & $23^{\rm h}32^{\rm m}55^{\rm s}.4$& $-03\degree11'34''$&  271 $\pm$ 6  &  336 $\pm$ 6  &  263 $\pm$ 8   \\ 

HELMS3 & $00^{\rm h}02^{\rm m}15^{\rm s}.9$& $-01\degree28'29''$&  643 $\pm$ 7  &  510 $\pm$ 6  &  258 $\pm$ 7   \\ 

HELMS4$^{\star,\dagger}$ & $00^{\rm h}44^{\rm m}10^{\rm s}.2$& $+01\degree18'21''$&  113 $\pm$ 7  &  177 $\pm$ 6  &  209 $\pm$ 8   \\ 

HELMS5$^{\ddagger}$ & $23^{\rm h}40^{\rm m}51^{\rm s}.5$& $-04\degree19'38''$&  151 $\pm$ 6  &  209 $\pm$ 6  &  205 $\pm$ 8   \\ 

HELMS6 & $23^{\rm h}36^{\rm m}20^{\rm s}.8$& $-06\degree08'28''$&  193 $\pm$ 7  &  252 $\pm$ 6  &  202 $\pm$ 8   \\ 

HELMS7$^{\ddagger}$ & $23^{\rm h}24^{\rm m}39^{\rm s}.5$& $-04\degree39'36''$&  214 $\pm$ 7  &  218 $\pm$ 7  &  172 $\pm$ 9   \\ 

HELMS8$^{\ddagger}$ & $00^{\rm h}47^{\rm m}14^{\rm s}.2$& $+03\degree24'54''$&  312 $\pm$ 6  &  244 $\pm$ 7  &  168 $\pm$ 8   \\

HELMS9$^{\ddagger}$ & $00^{\rm h}47^{\rm m}23^{\rm s}.6$& $+01\degree57'51''$&  398 $\pm$ 6  &  320 $\pm$ 6  &  164 $\pm$ 8   \\ 

HELMS10$^{\star}$ & $00^{\rm h}52^{\rm m}58^{\rm s}.6$& $+06\degree13'19''$&  88 $\pm$ 6  &  129 $\pm$ 6  &  155 $\pm$ 7   \\

HELMS11$^{\star,\dagger}$ & $00^{\rm h}39^{\rm m}29^{\rm s}.6$& $+00\degree24'26''$&  140 $\pm$ 7  &  157 $\pm$ 7  &  154 $\pm$ 8   \\

HELMS12 & $23^{\rm h}56^{\rm m}01^{\rm s}.5$& $-07\degree11'42''$&  178 $\pm$ 7  &  184 $\pm$ 6  &  154 $\pm$ 7   \\

HELMS13$^{\ddagger}$ & $00^{\rm h}16^{\rm m}15^{\rm s}.7$& $+03\degree24'35''$&  195 $\pm$ 6  &  221 $\pm$ 6  &  149 $\pm$ 7   \\

HELMS14 & $00^{\rm h}36^{\rm m}19^{\rm s}.8$& $+00\degree24'20''$&  251 $\pm$ 6  &  247 $\pm$ 6  &  148 $\pm$ 7   \\

HELMS15$^{\ddagger}$ & $23^{\rm h}32^{\rm m}55^{\rm s}.7$& $-05\degree34'26''$&  148 $\pm$ 6  &  187 $\pm$ 6  &  147 $\pm$ 9   \\

HELMS16 & $23^{\rm h}18^{\rm m}57^{\rm s}.2$& $-05\degree30'35''$&  143 $\pm$ 7  &  183 $\pm$ 7  &  146 $\pm$ 8   \\

HELMS17 & $23^{\rm h}25^{\rm m}58^{\rm s}.3$& $-04\degree45'25''$&  190 $\pm$ 6  &  189 $\pm$ 6  &  142 $\pm$ 8   \\

HELMS18$^{\ddagger}$ & $00^{\rm h}51^{\rm m}59^{\rm s}.5$& $+06\degree22'41''$&  166 $\pm$ 6  &  195 $\pm$ 6  &  135 $\pm$ 7   \\

HELMS19 & $23^{\rm h}22^{\rm m}10^{\rm s}.3$& $-03\degree35'59''$&  114 $\pm$ 6  &  160 $\pm$ 7  &  134 $\pm$ 8   \\

HELMS20 & $23^{\rm h}37^{\rm m}28^{\rm s}.8$& $-04\degree51'06''$&  162 $\pm$ 6  &  178 $\pm$ 7  &  132 $\pm$ 8   \\

HELMS21 & $00^{\rm h}18^{\rm m}00^{\rm s}.1$& $-06\degree02'35''$&  206 $\pm$ 6  &  186 $\pm$ 7  &  130 $\pm$ 7   \\

HELMS22$^{\ddagger}$ & $00^{\rm h}16^{\rm m}26^{\rm s}.0$& $+04\degree26'13''$&  117 $\pm$ 7  &  151 $\pm$ 6  &  127 $\pm$ 7   \\

HELMS23 & $00^{\rm h}58^{\rm m}41^{\rm s}.2$& $-01\degree11'49''$&  391 $\pm$ 7  &  273 $\pm$ 6  &  126 $\pm$ 8   \\

HELMS24$^{\star,\dagger}$ & $00^{\rm h}38^{\rm m}14^{\rm s}.1$& $-00\degree22'52''$&  82 $\pm$ 6  &  120 $\pm$ 6  &  126 $\pm$ 7   \\

HELMS25 & $00^{\rm h}41^{\rm m}24^{\rm s}.0$& $-01\degree03'07''$&  178 $\pm$ 6  &  186 $\pm$ 7  &  125 $\pm$ 8   \\

HELMS26$^{\star}$ & $00^{\rm h}47^{\rm m}47^{\rm s}.1$& $+06\degree14'44''$&  85 $\pm$ 7  &  119 $\pm$ 6  &  125 $\pm$ 8   \\

HELMS27 & $00^{\rm h}37^{\rm m}58^{\rm s}.0$& $-01\degree06'22''$&  125 $\pm$ 7  &  144 $\pm$ 6  &  124 $\pm$ 8   \\

HELMS28 & $00^{\rm h}30^{\rm m}09^{\rm s}.2$& $-02\degree06'25''$&  114 $\pm$ 6  &  135 $\pm$ 6  &  122 $\pm$ 7   \\

HELMS29$^{\star,\dagger}$ & $00^{\rm h}22^{\rm m}20^{\rm s}.9$& $-01\degree55'24''$&  66 $\pm$ 6  &  102 $\pm$ 6  &  121 $\pm$ 7   \\

HELMS30 & $00^{\rm h}10^{\rm m}27^{\rm s}.1$& $-02\degree46'24''$&  185 $\pm$ 6  &  170 $\pm$ 6  &  121 $\pm$ 7   \\

HELMS31 & $00^{\rm h}13^{\rm m}53^{\rm s}.5$& $-06\degree02'00''$&  178 $\pm$ 7  &  176 $\pm$ 6  &  120 $\pm$ 7   \\

HELMS32 &  $00^{\rm h}03^{\rm m}36^{\rm s}.9$& $+01\degree40'13''$&  103 $\pm$ 6  &  112 $\pm$ 6 &  119 $\pm$ 7   \\

HELMS33 & $00^{\rm h}30^{\rm m}32^{\rm s}.1$& $-02\degree11'53''$&  81 $\pm$ 7  &  98 $\pm$ 6  &  118 $\pm$ 8   \\

HELMS34 & $00^{\rm h}27^{\rm m}19^{\rm s}.5$& $+00\degree12'04''$&  248 $\pm$ 6  &  206 $\pm$ 7  &  116 $\pm$ 8   \\

HELMS35 & $23^{\rm h}25^{\rm m}00^{\rm s}.1$& $-00\degree56'43''$&  122 $\pm$ 6  &  132 $\pm$ 7  &  114 $\pm$ 8   \\

HELMS36 & $23^{\rm h}43^{\rm m}14^{\rm s}.0$& $+01\degree21'52''$&  115 $\pm$ 6  &  115 $\pm$ 6  &  113 $\pm$ 8   \\

HELMS37 & $01^{\rm h}08^{\rm m}01^{\rm s}.8$& $+05\degree32'01''$&  122 $\pm$ 6  &  120 $\pm$ 6  &  113 $\pm$ 7   \\

HELMS38 & $00^{\rm h}22^{\rm m}08^{\rm s}.1$& $+03\degree40'44''$&  190 $\pm$ 6  &  157 $\pm$ 6  &  113 $\pm$ 7   \\

HELMS39 & $00^{\rm h}29^{\rm m}36^{\rm s}.3$& $+02\degree07'10''$&  81 $\pm$ 6  &  107 $\pm$ 6  &  112 $\pm$ 7   \\

HELMS40$^{\ddagger}$ & $23^{\rm h}53^{\rm m}32^{\rm s}.0$& $+03\degree17'18''$&  102 $\pm$ 6  &  123 $\pm$ 7  &  111 $\pm$ 7   \\

HELMS41 & $23^{\rm h}36^{\rm m}33^{\rm s}.5$& $-03\degree21'19''$&  130 $\pm$ 6  &  131 $\pm$ 6  &  110 $\pm$ 7   \\

HELMS42 & $23^{\rm h}40^{\rm m}14^{\rm s}.6$& $-07\degree07'38''$&  158 $\pm$ 6  &  154 $\pm$ 6  &  110 $\pm$ 8   \\

HELMS43 & $23^{\rm h}34^{\rm m}20^{\rm s}.4$& $-00\degree34'58''$&  156 $\pm$ 7  &  141 $\pm$ 5  &  109 $\pm$ 8   \\

HELMS44 & $23^{\rm h}14^{\rm m}47^{\rm s}.5$& $-04\degree56'58''$&  220 $\pm$ 8  &  141 $\pm$ 7  &  106 $\pm$ 8   \\

HELMS45 & $00^{\rm h}12^{\rm m}26^{\rm s}.9$& $+02\degree08'10''$&  107 $\pm$ 6  &  142 $\pm$ 6  &  106 $\pm$ 7   \\

HELMS46 & $00^{\rm h}46^{\rm m}22^{\rm s}.3$& $+07\degree35'09''$&  82 $\pm$ 9  &  113 $\pm$ 9  &  105 $\pm$ 10   \\

HELMS47 & $23^{\rm h}49^{\rm m}51^{\rm s}.6$& $-03\degree00'19''$&  186 $\pm$ 7  &  167 $\pm$ 6  &  105 $\pm$ 8   \\

HELMS48 & $23^{\rm h}28^{\rm m}33^{\rm s}.6$& $-03\degree14'16''$&  49 $\pm$ 6  &  104 $\pm$ 6  &  105 $\pm$ 8   \\

HELMS49 & $23^{\rm h}37^{\rm m}21^{\rm s}.9$& $-06\degree47'40''$&  173 $\pm$ 6  &  161 $\pm$ 7  &  105 $\pm$ 8   \\

HELMS50 & $23^{\rm h}51^{\rm m}01^{\rm s}.7$& $-02\degree44'26''$&  112 $\pm$ 6  &  124 $\pm$ 6  &  105 $\pm$ 7   \\

HELMS51 & $23^{\rm h}26^{\rm m}17^{\rm s}.5$& $-02\degree53'19''$&  86 $\pm$ 6  &  109 $\pm$ 6  &  104 $\pm$ 7   \\

HELMS52 & $23^{\rm h}37^{\rm m}27^{\rm s}.1$& $-00\degree23'43''$&  182 $\pm$ 6  &  157 $\pm$ 6  &  104 $\pm$ 8   \\

HELMS53$^{\dagger}$ & $00^{\rm h}45^{\rm m}32^{\rm s}.6$& $-00\degree01'23''$&  48 $\pm$ 7  &  85 $\pm$ 6  &  103 $\pm$ 8   \\

HELMS54 & $00^{\rm h}27^{\rm m}18^{\rm s}.1$& $+02\degree39'43''$&  69 $\pm$ 6  &  87 $\pm$ 6  &  103 $\pm$ 7   \\

HELMS55 & $23^{\rm h}28^{\rm m}31^{\rm s}.8$& $-00\degree40'35''$&  95 $\pm$ 7  &  120 $\pm$ 6  &  102 $\pm$ 7   \\

HELMS56 & $00^{\rm h}13^{\rm m}25^{\rm s}.7$& $+04\degree25'09''$&  89 $\pm$ 6  &  98 $\pm$ 6  &  102 $\pm$ 7   \\

HELMS57 & $00^{\rm h}35^{\rm m}19^{\rm s}.7$& $+07\degree28'06''$&  134 $\pm$ 6  &  135 $\pm$ 7  &  101 $\pm$ 8   \\ 
\hline
\end{longtable*}

{\footnotesize \hspace{1.6 cm} $^{\star}$: Red source in \citet{Asboth2016}\\
\vspace{1 mm}
\hspace{1.8 cm} $^{\dagger}$: ACT identified lensed DSFG \citep{Su2015} \\
\vspace{1 mm}
\hspace{1.8 cm} $^{\ddagger}$: ALMA Cycle 1 targets (PI: Eales) \\}


\newpage




\begin{longtable*}{llllll}
\caption{HerS lens candidates ($S_{500}>100$\,mJy)}\\
\hline

Object ID & RA & Dec & $S_{250}$\,(mJy) & $S_{350}$\,(mJy) &
$S_{500}$\,(mJy) \\ 
\\
\hline
HERS1$^{\star,\dagger}$ & $02^{\rm h}09^{\rm m}41^{\rm s}.2$& $+00\degree15'58''$&   826 $\pm$ 7  &   912 $\pm$ 7  &   718 $\pm$ 8  \\

HERS2 & $01^{\rm h}20^{\rm m}41^{\rm s}.6$& $-00\degree27'05''$&   240 $\pm$ 6  &   260 $\pm$ 6  &   198 $\pm$ 7  \\

HERS3 & $01^{\rm h}27^{\rm m}54^{\rm s}.1$& $+00\degree49'40''$&   253 $\pm$ 6  &   250 $\pm$ 6  &   191 $\pm$ 7  \\

HERS4$^{\dagger}$ & $01^{\rm h}16^{\rm m}40^{\rm s}.1$& $-00\degree04'54''$&   137 $\pm$ 7  &   196 $\pm$ 7  &   190 $\pm$ 8  \\

HERS5 & $01^{\rm h}26^{\rm m}20^{\rm s}.5$& $+01\degree29'50''$&   268 $\pm$ 8  &   228 $\pm$ 7  &   133 $\pm$ 9  \\

HERS6 & $01^{\rm h}03^{\rm m}01^{\rm s}.2$& $-00\degree33'01''$&   121 $\pm$ 7  &   147 $\pm$ 6  &   130 $\pm$ 8  \\

HERS7 & $01^{\rm h}01^{\rm m}33^{\rm s}.8$& $+00\degree31'57''$&   165 $\pm$ 7  &   154 $\pm$ 6  &   122 $\pm$ 7  \\

HERS8 & $01^{\rm h}09^{\rm m}38^{\rm s}.9$& $-01\degree48'30''$&   146 $\pm$ 8  &   152 $\pm$ 7  &   118 $\pm$ 10  \\

HERS9 & $01^{\rm h}09^{\rm m}11^{\rm s}.7$& $-01\degree17'33''$&   393 $\pm$ 8  &   220 $\pm$ 8  &   118 $\pm$ 9  \\

HERS10 & $01^{\rm h}17^{\rm m}22^{\rm s}.3$& $+00\degree56'24''$&   105 $\pm$ 6  &   125 $\pm$ 6  &   117 $\pm$ 7  \\

HERS11 & $00^{\rm h}58^{\rm m}47^{\rm s}.3$& $-01\degree00'17''$&   63 $\pm$ 8  &   116 $\pm$ 7  &   115 $\pm$ 9  \\ 

HERS12 & $01^{\rm h}25^{\rm m}46^{\rm s}.3$& $-00\degree11'43''$&   152 $\pm$ 8  &   135 $\pm$ 7  &   114 $\pm$ 9  \\

HERS13 & $01^{\rm h}25^{\rm m}21^{\rm s}.0$& $+01\degree17'24''$&   165 $\pm$ 8  &   153 $\pm$ 7  &   114 $\pm$ 10  \\ 

HERS14 & $01^{\rm h}40^{\rm m}57^{\rm s}.3$& $-01\degree05'47''$&   136 $\pm$ 8  &   143 $\pm$ 8  &   112 $\pm$ 9  \\

HERS15 & $01^{\rm h}21^{\rm m}06^{\rm s}.9$& $+00\degree34'57''$&   94 $\pm$ 6  &   130 $\pm$ 7  &   110 $\pm$ 7  \\

HERS16 & $02^{\rm h}14^{\rm m}34^{\rm s}.4$& $+00\degree59'26''$&   110 $\pm$ 9  &   134 $\pm$ 8  &   109 $\pm$ 10  \\

HERS17 & $02^{\rm h}14^{\rm m}02^{\rm s}.6$& $-00\degree46'12''$&   110 $\pm$ 8  &   130 $\pm$ 8  &   105 $\pm$ 9  \\

HERS18 & $01^{\rm h}32^{\rm m}12^{\rm s}.2$& $+00\degree17'54''$&   176 $\pm$ 7  &   175 $\pm$ 6  &   104 $\pm$ 8  \\

HERS19 & $02^{\rm h}05^{\rm m}29^{\rm s}.1$& $+00\degree05'01''$&   89 $\pm$ 6  &   112 $\pm$ 6  &   102 $\pm$ 8  \\

HERS20 & $01^{\rm h}02^{\rm m}46^{\rm s}.1$& $+01\degree05'43''$&   107 $\pm$ 8  &   133 $\pm$ 8  &   102 $\pm$ 11  \\ 
\hline 
\end{longtable*}

{\footnotesize \hspace{1.6 cm} $^{\star}$: Lensed sources identified in \citet{Geach2015} \\
\vspace{1 mm}
\hspace{1.8 cm} $^{\dagger}$: ACT identified lensed DSFG \citep{Su2015} \\}





\begin{longtable*}{llll}
\caption{Properties of the foreground lensing galaxy and background lensed
  submillimeter galaxy.}\\
\hline

ID& Name& $z$ (foreground)$^1$  & $z$ (SMG)\\ 
\\
\hline
\endhead\\
HERS1& HERS J020941.1+001557 & 0.202$^2$ $\pm$ 0.00006&2.553$^3$
\\
HELMS2& HERMES J233255.5-031134 & 0.426 $\pm$ 0.1483 & 2.6899$^4$ 
\\
HELMS5& HERMES J234051.3-041937 & -- & 3.50$^5$ 
\\
HELMS6& HERMES J233620.7-060826 & 0.3958$^6$ $\pm$ 0.0007& 3.434$^5$ 
\\
HERS2& HERS J012041.5-002705 & 0.732 $\pm$ 0.0406& --
\\
HERS4& HERS J011640.0-000453 & 0.445 $\pm$ 0.0612& --
\\
HELMS7& HERMES J232439.4-043934 & -- & 2.473$^4$ 
\\
HELMS8& HERMES J004714.1+032453 & 0.478 $\pm$ 0.0847& 1.19$^5$ 
\\
HELMS9& HERMES J004723.3+015749 & 0.299 $\pm$ 0.0542& 1.441$^5$ 
\\
HELMS10& HERMES J005258.6+061317 & 0.241 $\pm$ 0.1176& --
\\
HELMS12& HERMES J235601.5-071144 & 0.775 $\pm$ 0.0800 & --
\\
HELMS13& HERMES J001615.8+032433 & 0.663$^2$ $\pm$ 0.00025& 2.765$^4$
\\
HELMS14& HERMES J003619.7+002420 & 0.258$^2$ $\pm$ 0.00005& --
\\
HELMS15& HERMES J233255.7-053424 & 0.976 $\pm$ 0.0565 & 2.4024$^4$ 
\\
HELMS18& HERMES J005159.4+062240 & -- & 2.392$^4$ 
\\
HELMS19& HERMES J232210.3-033600 & 0.143 $\pm$ 0.0869 & --
\\
HERS5& HERS J012620.5+012949 & 0.431 $\pm$ 0.0495& --
\\
HELMS21& HERMES J001800.1-060234 & 0.574 $\pm$ 0.1237& --
\\
HERS6& HERS J010301.2-003300 & 0.429$^2$ $\pm$ 0.00010& 2.2153$^4$
\\
HELMS22& HERMES J001626.0+042613 & 0.218 $\pm$ 0.0175& 2.5093$^4$ 
\\
HELMS23& HERMES J005841.0-011148 & 0.375 $\pm$ 0.0777& --
\\
HELMS24& HERMES J003813.9-002253 & 0.169 $\pm$ 0.0846& --
\\
HELMS25& HERMES J004123.8-010311 & 0.271 $\pm$ 0.0716& --
\\
HELMS28& HERMES J003009.2-020623 & 0.415 $\pm$ 0.1188& --
\\
HERS7& HERS J010133.7+003157 & 0.334 $\pm$ 0.1205 & --
\\
HELMS29& HERMES J002220.9-015520& -- & 5.162$^7$
\\
HELMS30& HERMES J001027.1-024625 & 0.851 $\pm$ 0.1137& --
\\
HELMS31& HERMES J001353.6-060200 & 0.604 $\pm$ 0.1777& --
\\
HERS8& HERS J010938.8-014829 & 0.378 $\pm$ 0.0821& --
\\
HERS9& HERS J010911.7-011732 & 0.853$^2$ $\pm$ 0.00008& --
\\
HERS10& HERS J011722.2+005624 & 0.873 $\pm$ 0.0531 & --
\\
HELMS34& HERMES J002719.6+001203 & 0.512 $\pm$ 0.1403 & --
\\
HERS11& HERS J005847.2-010016 & 0.717 $\pm$ 0.1537& --
\\
HELMS35& HERMES J232500.1-005644 & 0.299 $\pm$ 0.1217& --
\\
HERS12& HERS J012546.3-001143 & 0.893$^2$ $\pm$ 0.00039 & --
\\
HELMS36& HERMES J234314.0+012152 & 0.489 $\pm$ 0.0286& --
\\
HELMS38& HERMES J002207.9+034044 & 0.217$^2$ $\pm$ 0.00004& --
\\
HERS14& HERS J014057.3-010547 & 0.370 $\pm$ 0.0724& --
\\
HELMS39& HERMES J002936.3+020706 & 0.770 $\pm$ 0.1217& --
\\
HELMS40& HERMES J235331.7+031717 & 0.821 $\pm$ 0.0940& --
\\
HERS15& HERS J012106.8+003456 & 0.698 $\pm$ 0.0721& --
\\
HELMS42& HERMES J234014.4-070737 & 0.460 $\pm$ 0.0487& --
\\
HELMS43& HERMES J233420.2-003455 & 0.139 $\pm$ 0.0591 & --
\\
HERS16& HERS J021434.4+005925 & 0.492 $\pm$ 0.1321& --
\\
HELMS44& HERMES J231447.6-045657 & 0.451 $\pm$ 0.0816& --
\\
HELMS46& HERMES J004622.2+073514 & 0.442 $\pm$ 0.1621& --
\\
HERS17& HERS J021402.5-004612 & 0.369$^2$ $\pm$ 0.00012& --
\\
HELMS50& HERMES J235101.7-024425& 0.134$^2$ $\pm$ 0.00002& --
\\
HERS18& HERS J013212.2+001754 & 0.495 $\pm$ 0.0755& --
\\
HERS19& HERS J020529.0+000500 & 0.455 $\pm$ 0.0764& --
\\
HELMS56& HERMES J001325.8+042506 & 0.485 $\pm$ 0.1861& --
\\
HELMS57& HERMES J003519.7+072808& 0.094 $\pm$ 0.1104 & --
\\
\hline

\end{longtable*}

{\footnotesize
\hspace{1.1 cm} $^1$: SDSS DR12 PhotoZ KD tree method, unless otherwise noted \\
\vspace{0.7 mm}
\hspace{1.3 cm} $^2$: SDSS DR12 Spectroscopic Redshift \\
\vspace{0.7 mm}
\hspace{1.3 cm} $^3$: \citealp{Geach2015} \\
\vspace{0.7 mm}
\hspace{1.3 cm} $^4$: GBT/Zspectrometer (Harris et al., in prep), CARMA and PdBI (Riechers et al., in prep) \\
\vspace{0.7 mm}
\hspace{1.3 cm} $^5$: CARMA and PdBI (Riechers et al., in prep) \\
\vspace{0.7 mm}
\hspace{1.3 cm} $^6$: GTC/OSIRIS spectroscopic redshift, Marques-Chaves et al., in prep \\
\vspace{0.7 mm}
\hspace{1.3 cm} $^7$: \citet{Asboth2016} \\}



\begin{figure*}
\centering
\includegraphics[trim=0cm 0cm 8cm 0cm, scale=0.8]{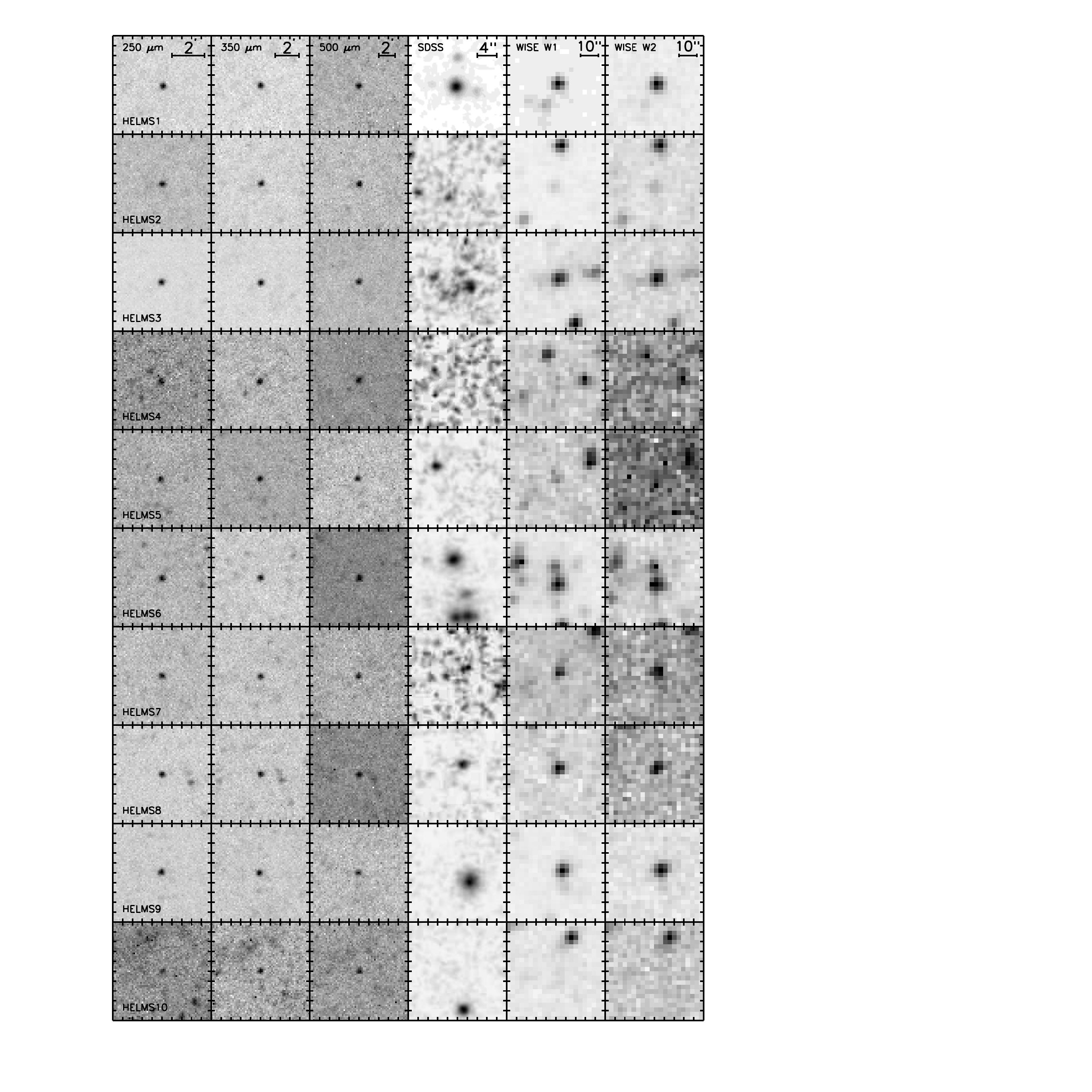}
\caption{The postage stamp images of the candidate gravitationally lensed DSFGs in the HeLMS field in the {\it Herschel} far-infrared, SDSS optical and Wide-field Infrared Survey Explorer (WISE; \citealp{Wright2010}) infrared bands. The {\it Herschel} cutouts are from our {\sc sanepic} maps (see \citealp{Asboth2016}) with the coordinates derived from the higher resolution 250\,$\mu m$ observations. The SDSS data are from deep co-adds by \citet{Jiang2014} when available and single epoch observations from SDSS DR12, when deeper data is not available. The infrared data are in WISE W1 (at 3.4\,$\mu m$) and W2 (at 4.6\,$\mu m$) bands using unWISE images \citep{Lang2014}. The image scales are shown for the first row. The lensed candidates are point-like bright targets in the {\it Herschel} SPIRE cutouts. The SDSS and WISE cutouts show the lensing foreground galaxies.}
\end{figure*}

\addtocounter{figure}{-1}

\begin{figure*}
\centering
\includegraphics[trim=0cm 0cm 8cm 0cm, scale=0.8]{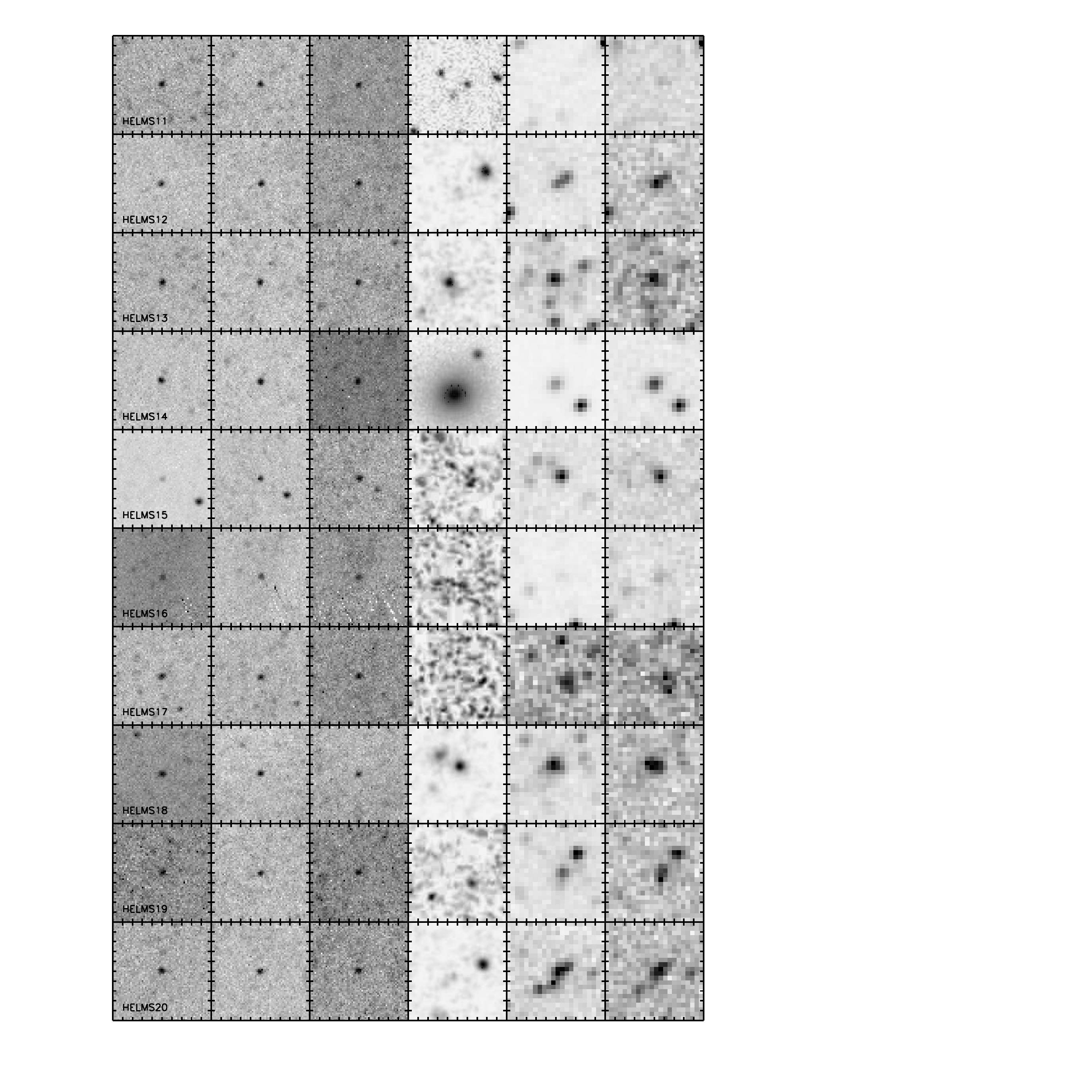}
\caption{Continued.}
\end{figure*}

\addtocounter{figure}{-1}

\begin{figure*}
\centering
\includegraphics[trim=0cm 0cm 8cm 0cm, scale=0.8]{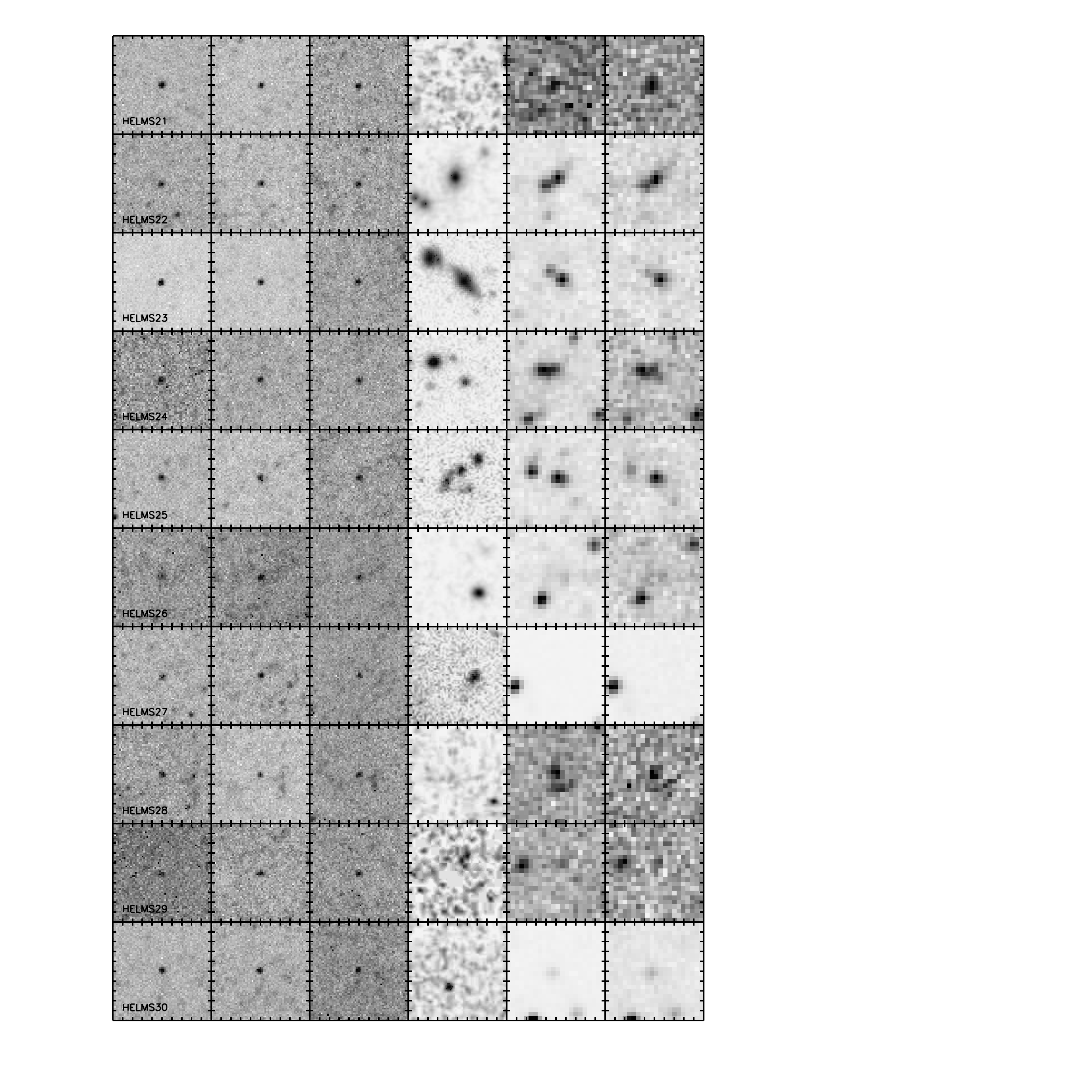}
\caption{Continued.}
\end{figure*}

\addtocounter{figure}{-1}

\begin{figure*}
\centering
\includegraphics[trim=0cm 0cm 8cm 0cm, scale=0.8]{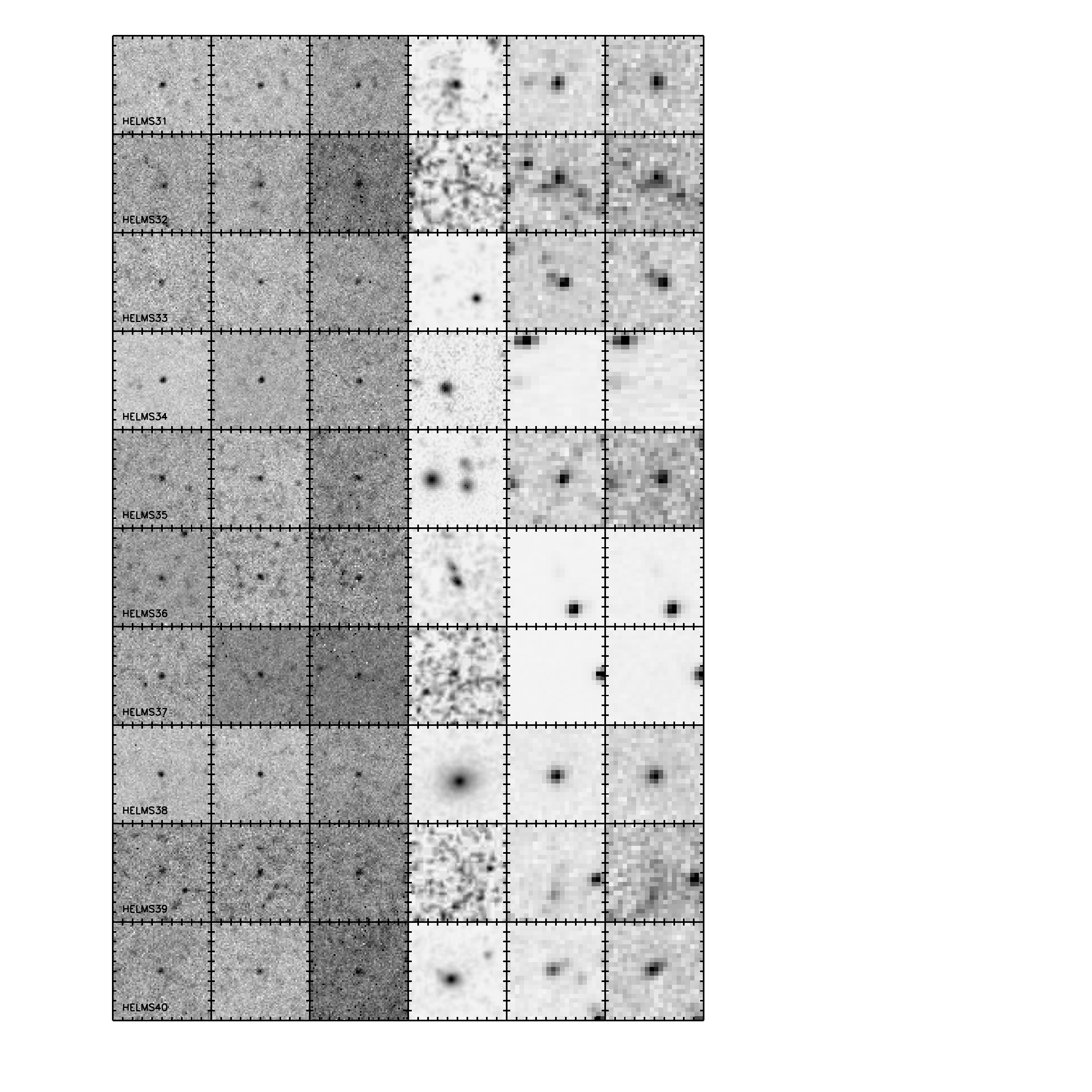}
\caption{Continued.}
\end{figure*}

\addtocounter{figure}{-1}

\begin{figure*}
\centering
\includegraphics[trim=0cm 0cm 8cm 0cm, scale=0.8]{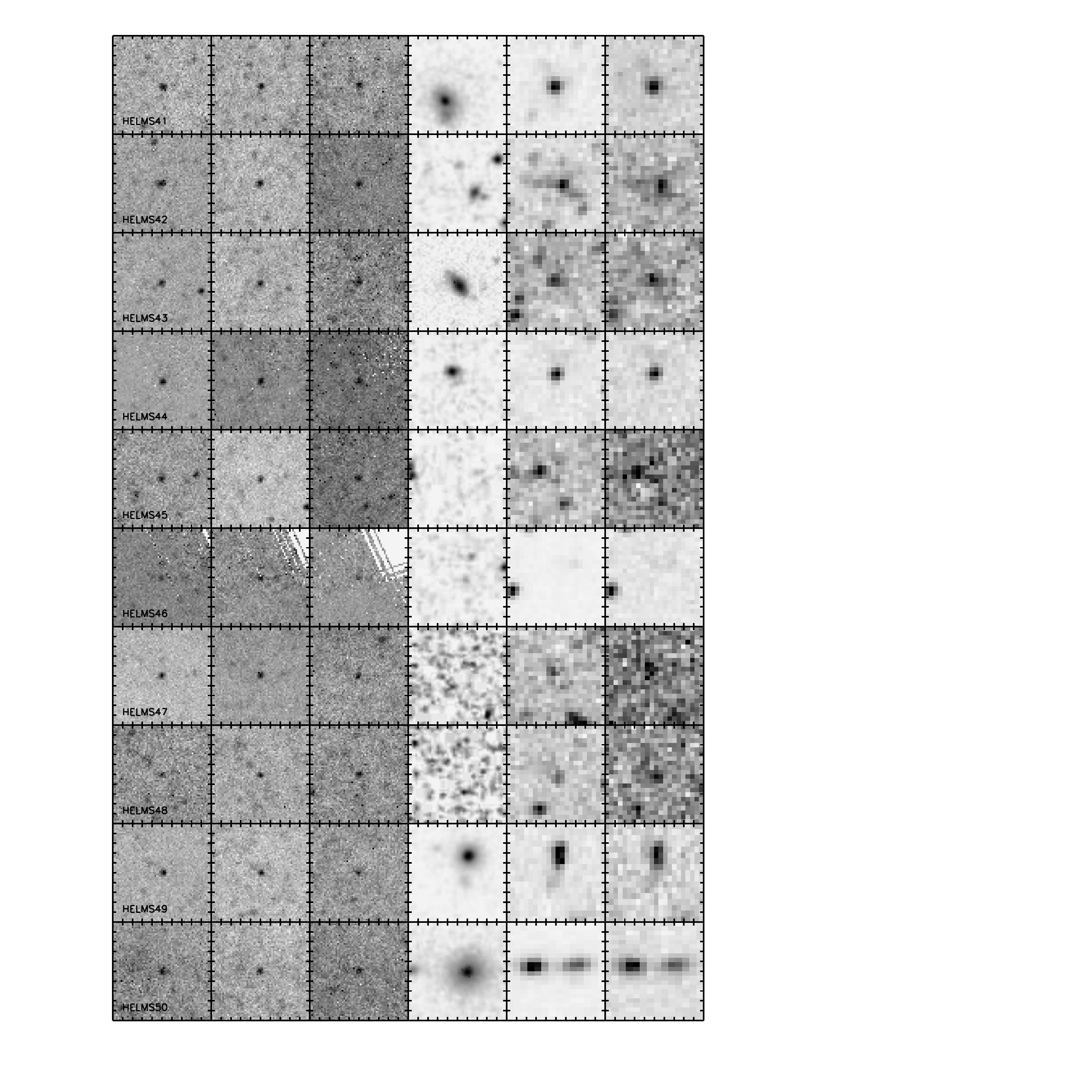}
\caption{Continued.}
\end{figure*}

\addtocounter{figure}{-1}

\begin{figure*}
\centering
\includegraphics[trim=0cm 0cm 8cm 0cm, scale=0.8]{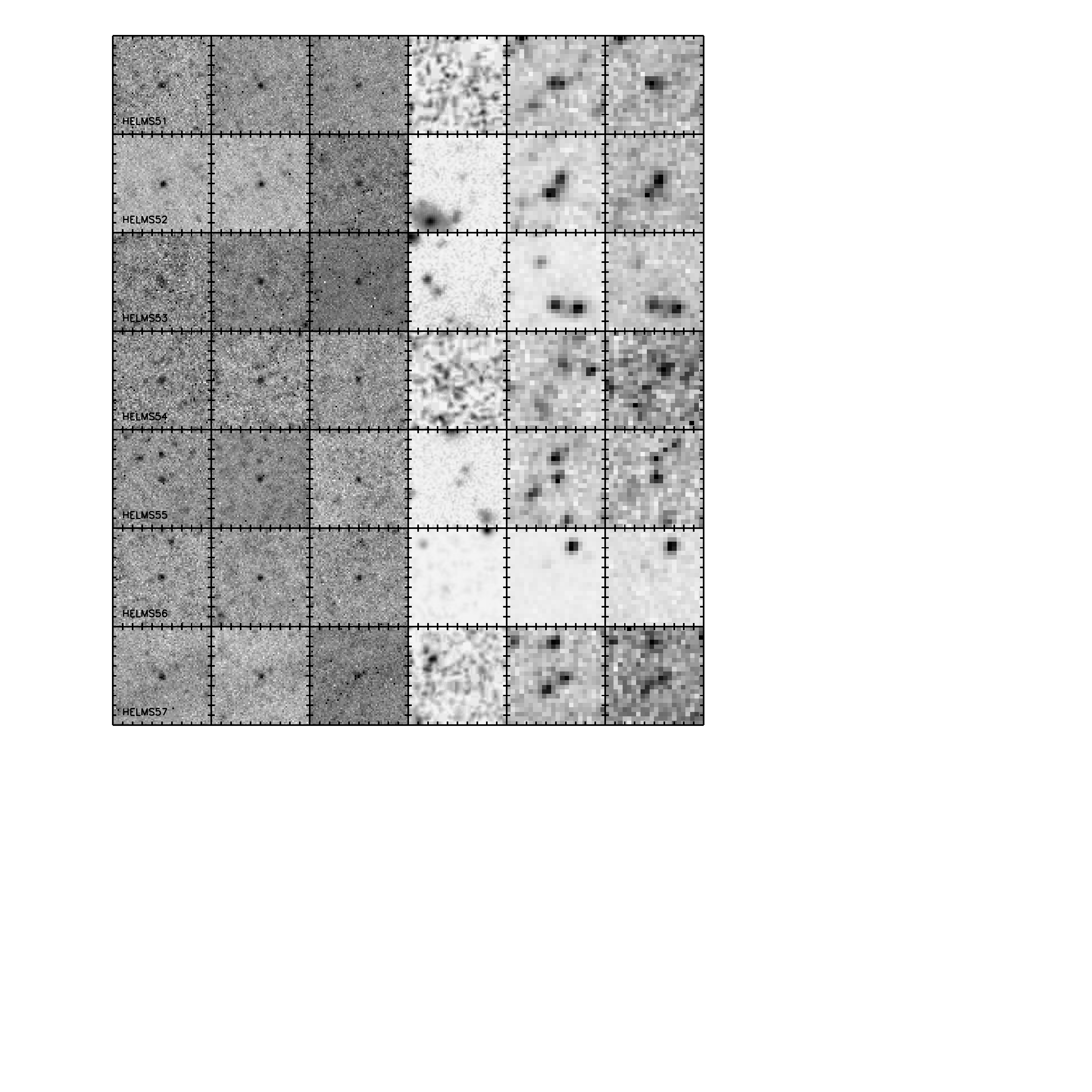}
\caption{Continued.}
\end{figure*}

\begin{figure*}
\centering
\includegraphics[trim=0cm 0cm 8cm 0cm, scale=0.8]{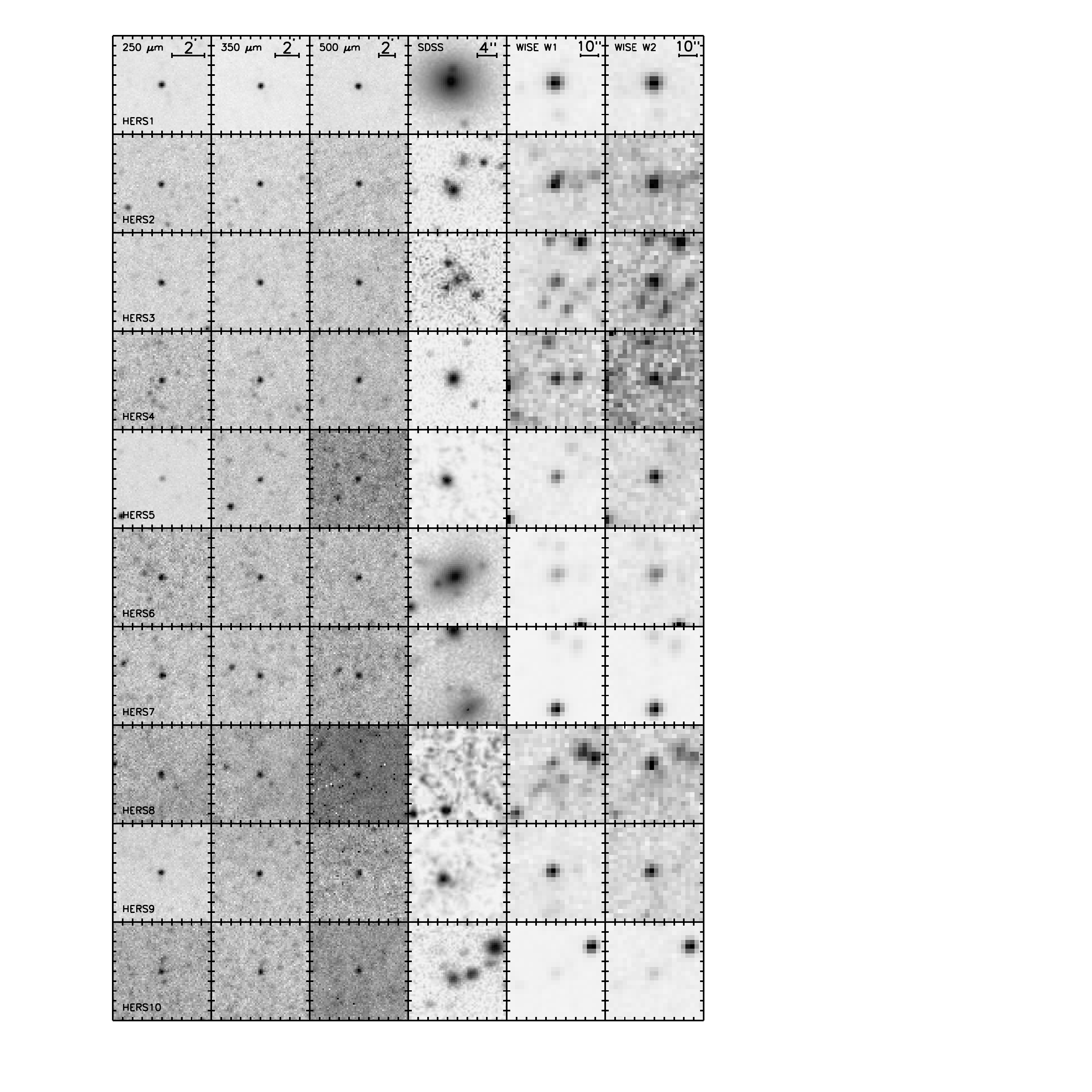}
\caption{Same as Figure 9 but for the HerS sample of gravitationally lensed DSFG candidates.}
\end{figure*}

\addtocounter{figure}{-1}

\begin{figure*}
\centering
\includegraphics[trim=0cm 0cm 8cm 0cm, scale=0.8]{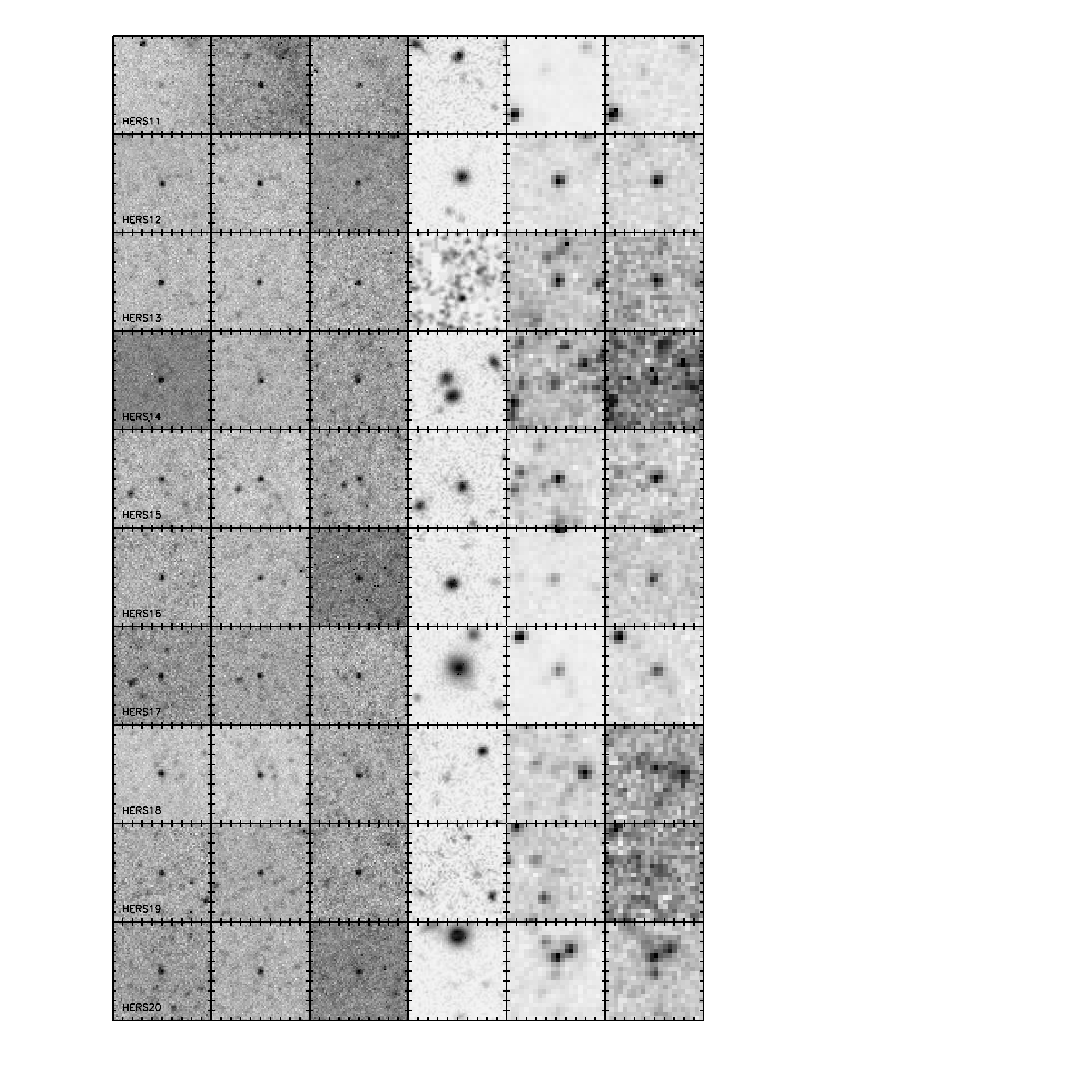}
\caption{Continued.}
\end{figure*}

\end{document}